\documentclass{emulateapj}
\usepackage{gensymb}
\usepackage{upgreek}
\usepackage{hyperref}

\begin{document}

\shortauthors{Esplin \& Luhman}
\shorttitle{Census of Corona Australis}

\title{A Census of Stars and Disks in Corona Australis\altaffilmark{1}}

\author{
T. L. Esplin\altaffilmark{2}
and
K. L. Luhman\altaffilmark{3,4}}

\altaffiltext{1}
{Based on observations made with the Gaia mission, the Two Micron All Sky
Survey, the Wide-field Infrared Survey Explorer, the Spitzer Space Telescope,
the NASA Infrared Telescope Facility, Cerro Tololo Inter-American Observatory, 
and Magellan Observatory.}
\altaffiltext{2}{Steward Observatory, University of Arizona, Tucson, AZ, 85719,
USA; taranesplin@email.arizona.edu}
\altaffiltext{3}{Department of Astronomy and Astrophysics, The Pennsylvania
State University, University Park, PA 16802; taran.esplin@psu.edu.}
\altaffiltext{4}{Center for Exoplanets and Habitable Worlds,
The Pennsylvania State University, University Park, PA 16802.}

\begin{abstract}

We have performed a census of the young stellar populations near
the Corona Australis molecular cloud using photometric and kinematic data
from several sources, particularly Gaia EDR3, and spectroscopy of hundreds
of candidate members. 
We have compiled a catalog of 393 members of Corona Australis,
(39 at $>$M6), 293 (36) of which are spectroscopically classified 
for the first time in this work.  
We find that Corona Australis can be described in terms of
two stellar populations, a younger one (few Myr) that is
partially embedded in the cloud (the Coronet Cluster) and an older
one ($\sim15$~Myr) that surrounds and extends beyond the cloud
(Upper Corona Australis).  
These populations exhibit similar space velocities, and we find no
evidence for distinct kinematic populations in Corona Australis, in
contrast to a recent study based on Gaia DR2.
The distribution of spectral types in Corona Australis reaches a maximum
at M5 ($\sim0.15$~$M_\odot$), indicating that the IMF 
has a similar characteristic mass as other nearby star-forming regions.
Finally, we have compiled mid-infrared photometry from 
the Wide-field Infrared Survey Explorer and the Spitzer Space Telescope
for the members of Corona Australis and we have used those data to identify
and classify their circumstellar disks. Excesses are detected for 122 stars,
a third of which are reported for the first time in this work.

\end{abstract}


\section{Introduction}
\label{sec:intro}

Identifying a well-defined sample of the members of a young stellar association
is required to measure many of its fundamental properties, which can
provide insight into the processes of star and planet formation.
A few such measurements include the distribution of masses of the stellar 
and substellar members (i.e., the initial mass function; IMF) and the 
frequency of disk-bearing members.
The minimum mass of the former can be used to test theories of star formation 
\citep[][references therein]{whi07} and the latter constrains the time 
available for giant planet formation \citep[e.g.,][]{mam09}.

Corona Australis is among the nearest star-forming regions
\citep[$\sim$149~pc,][]{neu08,gal20}, making it feasible to perform
a census of its members down to planetary masses ($\lesssim15$~M$_{\rm Jup}$). 
Based on early surveys,
the region appeared to consist of a young embedded population, the Coronet 
Cluster \citep{mey09,nis05,sic11}
and a less obscured distribution of stars surrounding the molecular cloud 
that may have older ages \citep{pet11,caz19}.
The high-precision astrometry from the Gaia mission \citep{per01,gaia16b}
has recently provided new measurements of the kinematic structure of
Corona Australis.
In the first data release of Gaia, \citet{gag18} discovered a 
group of 10 stars between Corona Australis and the Scorpius-Centaurus OB
association and near the distance of both regions, which they named Upper 
Corona Australis. Using the second data release of Gaia \citep[DR2,][]{gaia18}, 
\cite{gal20} identified many more candidate members of that new population
and found that it is kinematically distinct from and slightly older than
the stars near the Corona Australis molecular cloud.
However, the characterization of these stellar populations 
has been limited by the fact that only a small fraction of the candidate
members have been observed with spectroscopy, which is necessary for measuring
spectral types and confirming youth.

We have performed a census of the members of the stellar populations
near Corona Australis using astrometry, photometry, and spectroscopy
from a variety of sources. We begin by characterizing the kinematics of the
young stars projected against the Corona Australis 
cloud using the early installment of the third data release of Gaia
\citep[EDR3;][]{gaia21} (Section~\ref{sec:kin}). We compile objects
that have confirmation of youth and that satisfy our kinematic
criteria for membership (Section~\ref{sec:cat}) and we
identify new candidate members via their proper motions, parallaxes,
and photometry (Section~\ref{sec:cand}).
We present optical and infrared (IR) spectroscopy for a large number
of candidates to measure their spectral types and assess their youth
(Section~\ref{sec:spec}).
For stars that are adopted as members of Corona Australis, 
we characterize their space velocities, relative ages, and IMF and we 
check for the presence of circumstellar disks (Section~\ref{sec:stellarpop}).

\section{Kinematics of Young Stars Toward Corona Australis}
\label{sec:kin}

\subsection{Kinematic Populations from \citet{gal20}}

The proper motions and parallaxes for young stars toward
Corona Australis have been examined previously by \cite{gal20}.
In that study, candidate members of the region were identified using a 
Gaussian mixture model (GMM) that was applied to proper motions and
parallaxes ($\mu_\alpha$, $\mu_\delta$, $\pi$) from Gaia DR2
and a second statistical model that was applied to the Gaia photometry.
The Bayesian information criterion (BIC) favored a two component model,
which was interpreted as evidence that Corona Australis
was composed of two kinematically distinct populations. 
Most members of two populations had $b\lesssim-16\arcdeg$ and
$b\gtrsim-16\arcdeg$, which they referred to as on- and off-cloud, respectively.

\begin{figure}[h]
  \centering
  \includegraphics[trim = 0mm 0mm 0mm 0mm, clip=true, scale=.75]{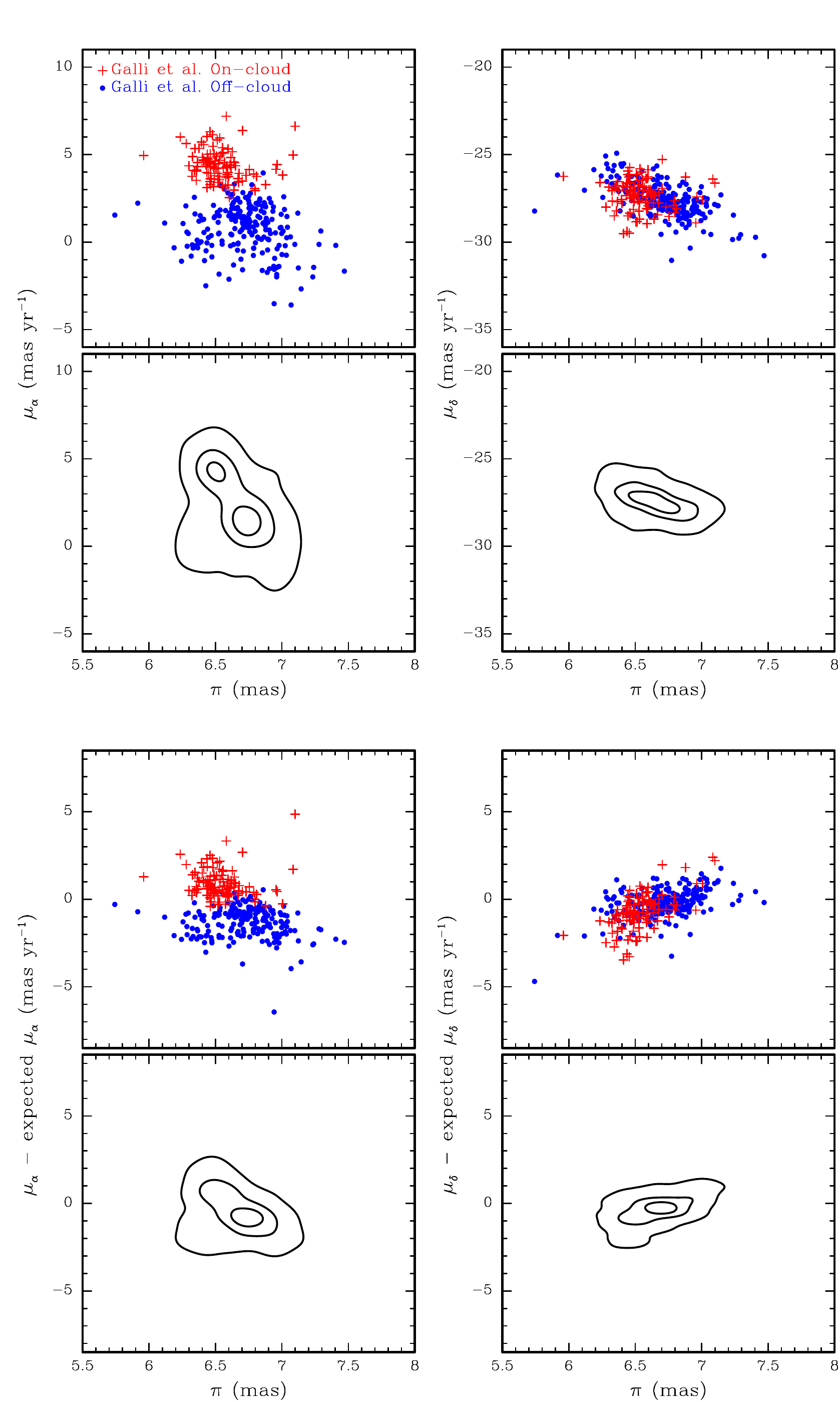}
\caption{
Proper motions (top two rows) and proper motion offsets (bottom two rows)
versus parallax based on Gaia EDR3 for two samples of candidate members of 
Corona Australis from \cite{gal20}.
The offsets are computed relative to the proper motions expected for the 
positions and parallaxes assuming the median space velocity of known members.
The contour levels are set at 10, 50, and 80\% of the maximum densities.
The bimodal nature of the candidates in $\mu_\alpha$ is a reflection of
projection effects, which are minimized in the proper motion offsets.
}
\label{fig:gaiagalli}
\end{figure}

The proper motion that corresponds to a given space
velocity varies with position on the sky due to projection effects.
To investigate the influence of projection effects on the results of 
\cite{gal20}, we have plotted the two populations from that study
in Figure~\ref{fig:gaiagalli} on diagrams of proper motions versus
parallax and proper motion offsets ($\Delta \mu_\alpha$, $\Delta \mu_\delta$)
versus parallax using data from Gaia EDR3. We have included density contours 
for each set of data. Projection effects are reduced in proper motion offsets, 
which are defined as the difference between the observed
proper motion and the motion expected at a given source's celestial
coordinates and parallax if it has a specified space velocity
\citep{luh18,esp19,esp20,luh20}. We have calculated the offsets assuming 
$U$, $V$, $W$, = $-6$,$-17$,$-8$~km~s$^{-1}$, which was the median velocity 
for a preliminary sample of members prior to our spectroscopic survey.
The median velocity of our final sample of members is
$-$3.9, $-$17.4, and $-$9.3~km~s$^{-1}$ (Section \ref{sec:uvw}). 
If the proper motion offsets were calculated using the latter velocity,
they would shift by a value that is very small and nearly constant for
all members, and thus would have no effect on our results.

\begin{figure}[h]
  \centering
  \includegraphics[trim = 0mm 0mm 0mm 0mm, clip=true, scale=.75]{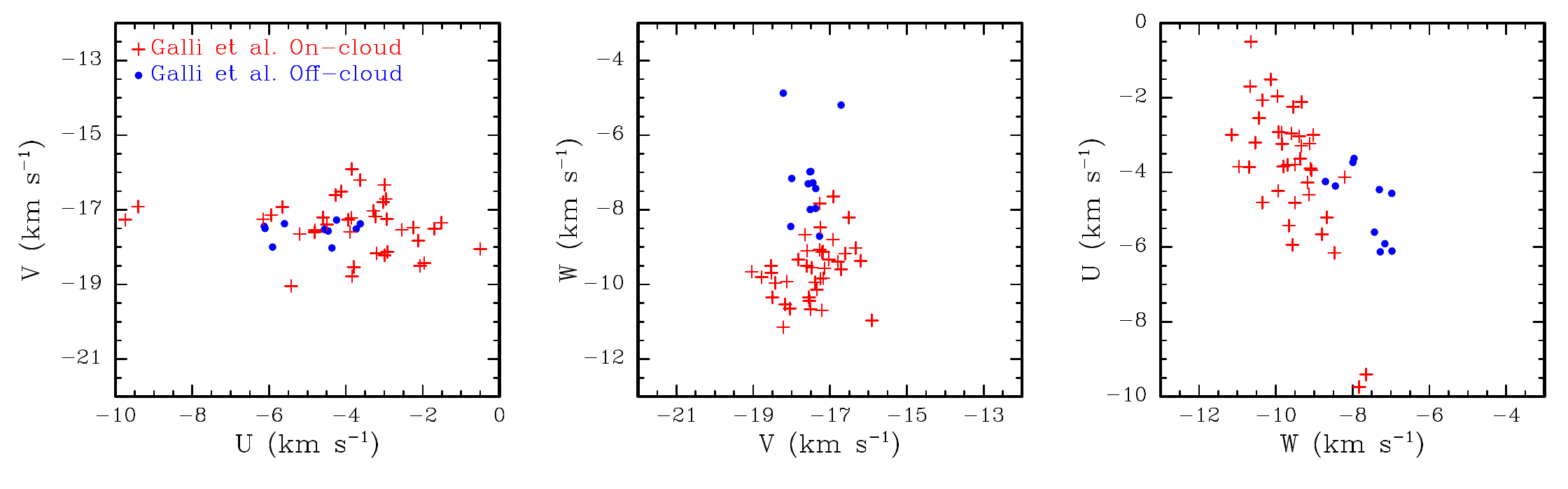}
\caption{
Space velocities for two samples of candidate members of
Corona Australis from \cite{gal20} that have measurements of
parallaxes and radial velocities.}
\label{fig:uvwgalli}
\end{figure}

In Figure~\ref{fig:gaiagalli}, two maxima are present in $\mu_\alpha$ 
versus $\pi$, which correspond to the two populations from \citet{gal20}.  
However, the candidates collapse into a single cluster in
$\Delta \mu_\alpha$ versus $\pi$, where projection effects are minimized.
Projection effects across Corona Australis lead to a broadening primarily
in $\mu_\alpha$, and a minimum in the surface density of the members 
near $b\sim-16\arcdeg$ results in two maxima in $\mu_\alpha$ instead 
of a single elongated distribution, and hence the identification of
two apparent populations by \citet{gal20}.
Meanwhile, in $\Delta \mu_\alpha$ versus $\pi$, the two populations from 
\citet{gal20} comprise two halves of the single cluster, which is a due 
to the fact that $\Delta \mu_\alpha$ is correlated with Galactic latitude 
among these stars and the two populations can be largely divided by latitude. 
As an additional comparison of the kinematics of the
two samples from \citet{gal20}, we have calculated the $UVW$ velocities for
the stars that have measurements of parallaxes and radial velocities
(Section~\ref{sec:uvw}). The resulting velocities are shown in 
Figure~\ref{fig:uvwgalli}.  In each pair of velocities, the two samples 
together form a single cluster, and there is no evidence of separate 
populations. The two samples do have systematically different velocities 
in $W$, which is a reflection of the different latitudes for the two samples 
and the correlation between $W$ and latitude among these stars (also 
manifested in the correlation between $\Delta \mu_\alpha$ and latitude).

\subsection{Corona Australis in Gaia EDR3}
\label{sec:edr3}

\begin{figure}[h]
\centering
\includegraphics[trim = 0mm 0mm 0mm 0mm, clip=true, scale=.8]{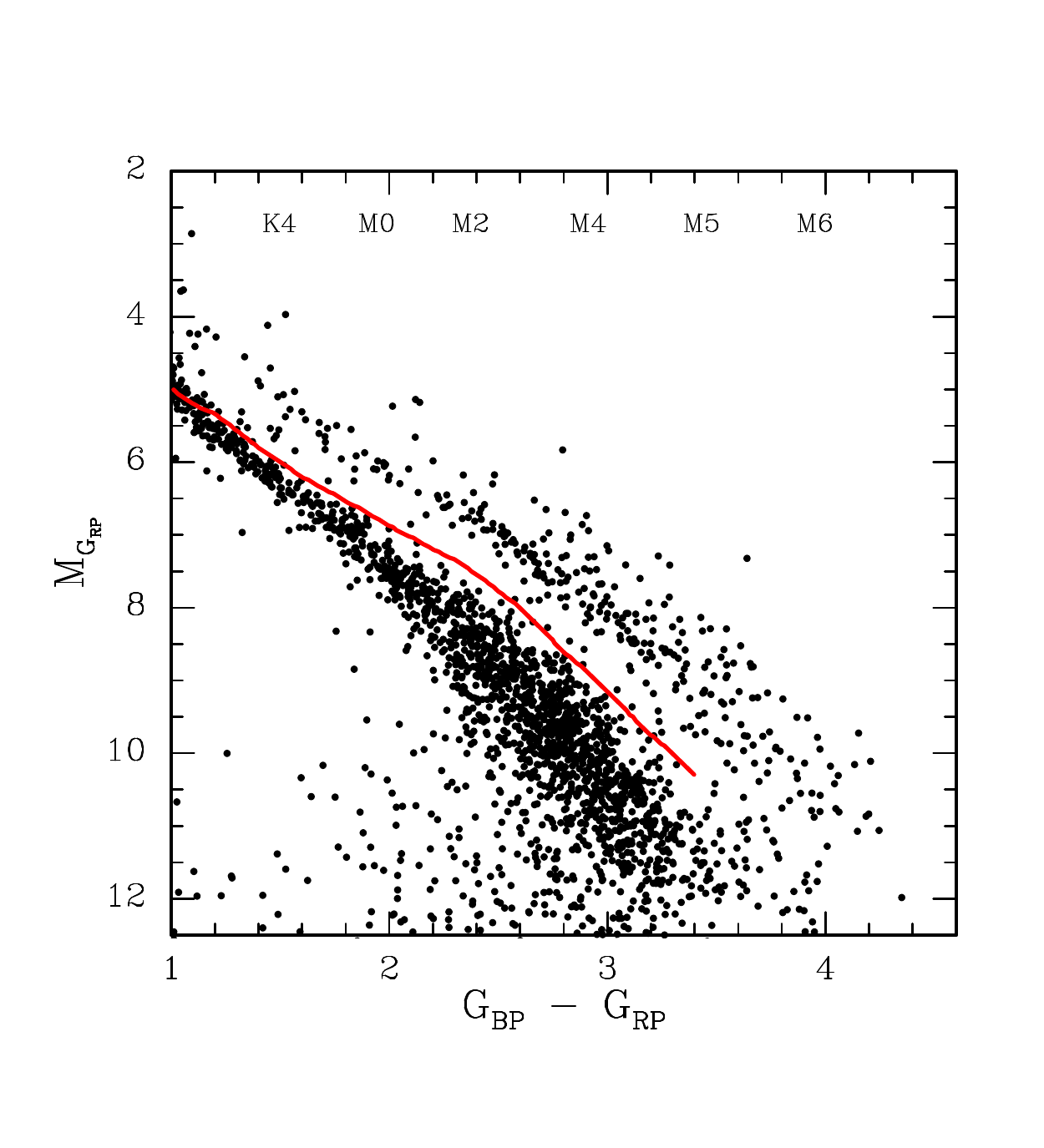}
\caption{
$M_{G_{\rm RP}}$ versus $G_{\rm BP}-G_{\rm RP}$ for Gaia EDR3 sources
near Corona Australis with $5.5<\pi<8.5$~mas.
The single star sequence of the Tuc-Hor association is indicated
\citep[45~Myr,][]{bel15} (solid line). The spectral types that correspond 
to these colors for young stars are marked \citep{luh21}.
}
\label{fig:gaiacmd}
\end{figure}

We have used data from Gaia EDR3 to characterize the kinematics of young
stars near the Corona Australis cloud.
We have selected sources from Gaia EDR3 with
right ascensions between 275 and 288.5$\arcdeg$,
declinations between $-$40 and $-32.5\arcdeg$,
parallaxes of 5.5--8.5~mas, $\sigma_\pi / \pi \leq 0.1$, and
renormalized unit weight errors $>1.6$ \citep[RUWEs,][]{lin18}.
The resulting sample is plotted in a diagram of $M_{G_{\rm RP}}$ versus 
$G_{\rm BP}-G_{\rm RP}$ in Figure~\ref{fig:gaiacmd}.
These bands cover 3300--6800~\AA\ ($G_{\rm BP}$) and 
6300--10500~\AA\ ($G_{\rm RP}$). 
The third Gaia band spans from 3300--10500~\AA\ ($G$).
We have included the single star sequence for the Tuc-Hor association
\citep[45~Myr,][]{bel15} from \cite{luh20}.
We have selected the sources appearing above that sequence
between $G_{BP}-G_{RP}=1.4$--3.4 as candidates for young low-mass stars.

\begin{figure}[h]
  \centering
  \includegraphics[trim = 0mm 0mm 0mm 0mm, clip=true, scale=.6]{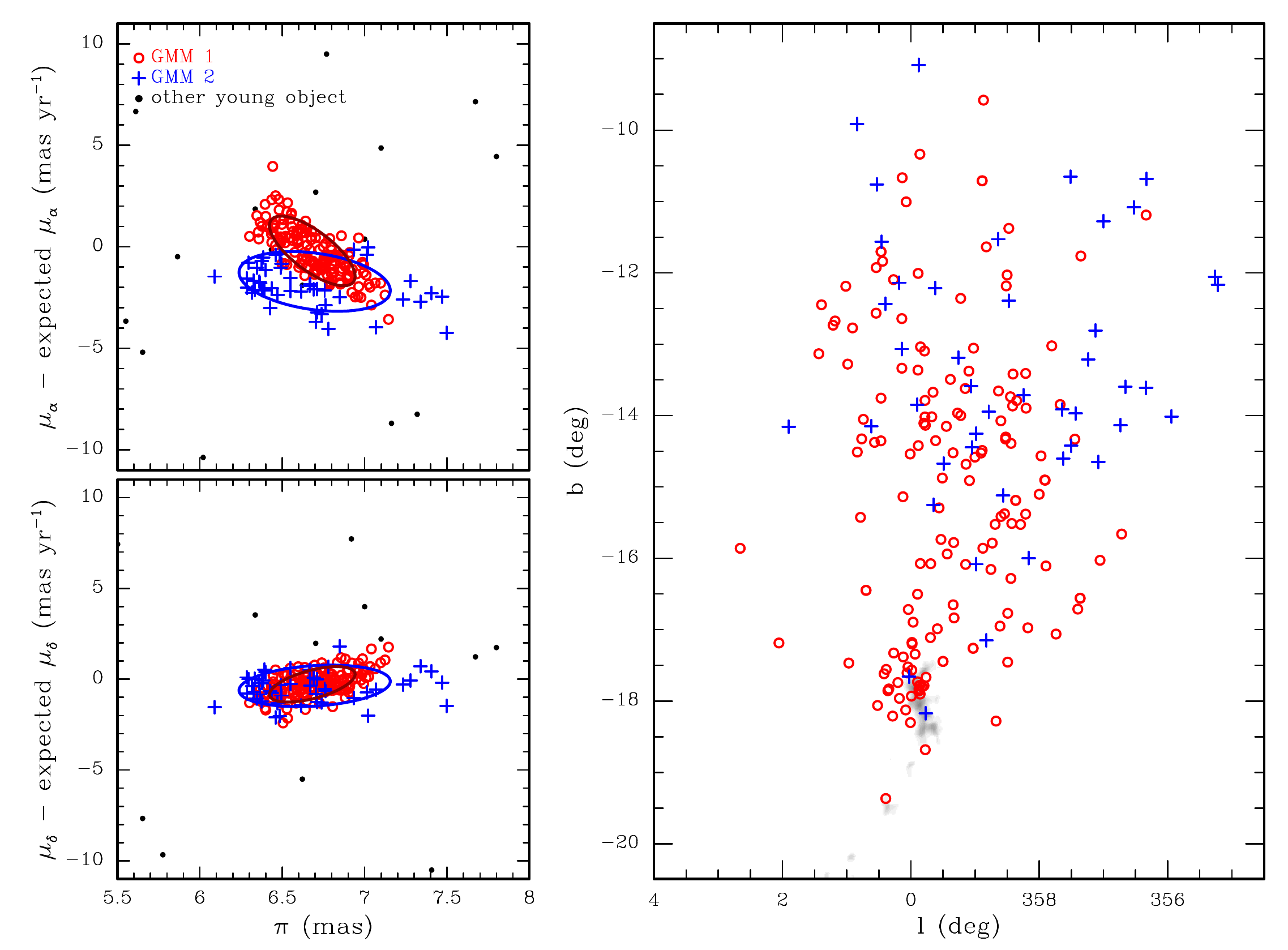}
\caption{
Proper motion offsets versus parallax for candidate young low-mass stars 
near Corona Australis from Figure~\ref{fig:gaiacmd} (left).
We have applied a GMM to these data, which consists of two clustered
components (red circles and blue pluses) and a noise population (black points).
The 2~$\sigma$ ellipses for the two model components are shown. 
The spatial distribution of the members of the components are plotted in
the right panel. The large overlap of the two components 
suggests that they comprise a single non-gaussian population rather than 
two distinct populations. Extinction from the Corona Australis cloud is 
displayed in the diagram on the right \citep[gray scale;][]{dob13}.
}
\label{fig:gaiapm}
\end{figure}

In the left panel of Figure~\ref{fig:gaiapm}, we have plotted 
$\Delta \mu_\alpha$ and $\Delta \mu_\delta$ versus $\pi$ for the
CMD-selected candidates that have $|\Delta \mu| < 10$~mas~yr$^{-1}$. 
Most of the candidates are tightly clustered in those diagrams,
which correspond to members of Corona Australis.
To separate the members from the small number of unclustered field stars,
we have applied GMMs to the proper motion offsets and parallaxes
using the {\tt mclust} library in R \citep{scr16,rcore}.
The models consist of a noise component for the field stars and
1--4 Gaussian components.
The noise component was initialized using the nearest-neighbor 
cleaning method from \cite{bye98} implemented in the {\tt prabcus} 
library \citep{chr19}. We found that two components (plus noise)
were statistically preferred according to the BIC. 
However, the two components are indistinguishable in terms of their
spatial distributions (Figure~\ref{fig:gaiapm}), ages, and extinctions
(Section~\ref{sec:age}), which
suggests that they do not comprise physically distinct populations. 
Meanwhile, the two components overlap to a large extent in the kinematic
parameters that were fit by the GMM (Figure~\ref{fig:gaiapm}), and thus could
easily represent a single kinematic population that has non-Gaussian
distributions in proper motion offsets and parallax. Such non-Gaussian
distributions have been observed for populations in other star-forming regions
and associations \citep{luh20}.

\section{Catalog of Known Members of Corona Australis}
\label{sec:cat}

\begin{figure}[h]
  \centering
  \includegraphics[trim = 0mm 0mm 0mm 0mm, clip=true, scale=.8]{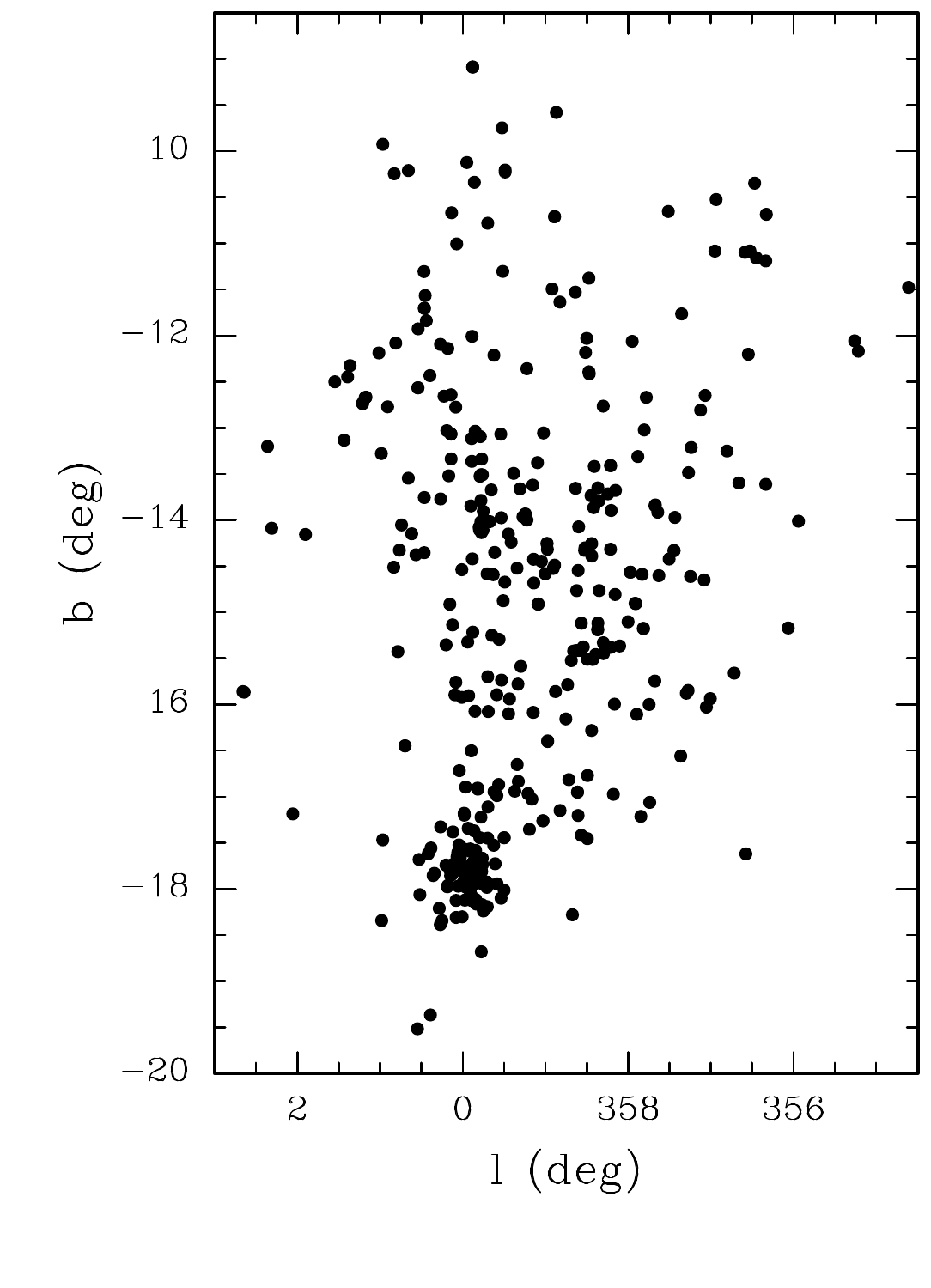}
\caption{
Spatial distribution of the adopted members of Corona Australis 
(Table~\ref{tab:mem}).
}
\label{fig:newmembers}
\end{figure}

We have compiled stars within the survey field encompassing
Corona Australis from Section~\ref{sec:edr3} that have evidence
of youth from previous studies or this work (Section~\ref{sec:cand})
and that are not rejected as non-members
by available measurements of proper motions and parallaxes.
We have identified stars that have evidence of youth,
parallaxes from Gaia EDR3 with errors less than 0.5~mas,
and $\Delta \mu_\alpha$, $\Delta \mu_\delta$, and $\pi$ that overlap
at 1~$\sigma$ with the 2~$\sigma$ ellipses of the GMM components in 
Figure~\ref{fig:gaiapm}, which results in 347 objects that are adopted
as members. We also have assigned membership to 13 stars that have
$\sigma_\pi<0.5$~mas and do not satisfy the preceding kinematic criteria, 
consisting of two objects that are candidate companions to adopted members 
(Gaia EDR3 6730717024217228928, 2MASS J19104337$-$3659092) and nine sources 
that may have unreliable astrometry based on values of RUWE greater than 1.6.
In addition, we have adopted 33 stars that lack parallaxes with 
$\sigma_\pi<0.5$~mas and have other data that support membership,
many of which are protostars \citep{nut05}.
The combined catalog of 393 adopted members is presented in Table~\ref{tab:mem}.
Previous evidence of youth (primarily from spectra) is available for 100 stars
and the remaining sources are spectroscopically classified for the first time 
in this work.
The spatial distribution of the adopted members is illustrated in a map
of their Galactic coordinates in Figure~\ref{fig:newmembers}.

Table~\ref{tab:mem} contains source designations from 
the AllWISE Source Catalog of the Wide-field Infrared Survey Explorer 
\citep[WISE,][]{wri10} and the Point Source Catalog of the Two Micron All Sky 
Survey \citep[2MASS,][]{skr06}; additional names from the literature;
spectral types from previous studies and this work;
astrometry from Gaia EDR3; if $\sigma_\pi<0.5$~mas, a flag indicating whether 
the Gaia kinematic criteria for membership are satisfied;
distance estimate based on the Gaia EDR3 parallax \citep{bai21};
proper motions measured from IR images in Section~\ref{sec:iracpm};
the most accurate available radial velocity measurement 
that has an error less than 4~km~s$^{-1}$;
$UVW$ velocities calculated from the radial velocity, proper motion,
and parallactic distance; photometry in bands from Gaia EDR3, 2MASS, WISE, 
and the Spitzer Space Telescope;
flags indicating whether excesses are detected in WISE and Spitzer bands;
disk classification if excess emission is detected; and extinction estimates.
The latter were derived from our near-IR spectra or $J-H$ and $J-K_s$ colors,
as indicated in Table~\ref{tab:mem}.

\section{Identification of Candidate Members}
\label{sec:cand}

\subsection{Proper Motions and Parallaxes from Gaia EDR3}
\label{sec:pp}

In Section~\ref{sec:edr3}, we characterized the proper motion offsets
and parallaxes for Corona Australis using candidate young low-mass stars 
selected from a CMD. We now use those kinematics to define criteria for
searching for candidate
members of Corona Australis at all magnitudes and colors in Gaia EDR3.
As in our compilation of adopted members in Section~\ref{sec:cat}, 
we consider sources from EDR3 that have locations within the survey
field from Section~\ref{sec:edr3}, $\sigma_\pi<0.5$~mas, and 
$\Delta \mu_\alpha$, $\Delta \mu_\delta$, and $\pi$ that overlap at 1~$\sigma$ 
with the 2~$\sigma$ ellipses of the GMM components in Figure~\ref{fig:gaiapm}. 
These criteria produce 346 objects that lack previous evidence of youth
or spectral classifications, 311 of which have RUWE$<$1.6.
We recover all but nine of the 254 candidates identified by 
\cite{gal20}, which consist of Gaia DR2 6728069984365602944,
6734711957980514432, 6736763921557552000, 6733973361050980352,
6733584275683948032, 6729972895366111232, 6730588072134320512,
6735323118961200000, and 6730960218154606464.

\subsection{Proper Motions from IR Imaging}
\label{sec:iracpm}

To search for candidate members of Corona Australis that are
too faint for Gaia parallax measurements,
such as heavily embedded stars or low-mass brown dwarfs,
we have measured proper motions of sources projected 
against the molecular cloud using multi-epoch IR imaging.
The first set of data that we have employed was taken with the Infrared 
Array Camera on the Spitzer Space Telescope \citep[IRAC;][]{faz04,wer04}.
IRAC operated in bands centered near 3.6, 4.5, 5.8, and 8.0~$\mu$m
([3.6], [4.5], [5.8], [8.0]]) during its cryogenic phase, which was from
August 2003 through May 2009.
It then collected data in only the first two bands for the remainder
of the mission.
The images from IRAC had a plate scale of $1\farcs2$ pixel$^{-1}$ and 
a field of view of $5\farcm2\times5\farcm2$.  The FWHM of point sources in 
[3.6]--[8.0] was $1\farcs6$--$1\farcs9$.

\begin{figure}[h]
  \centering
  \includegraphics[trim = 0mm 0mm 0mm 0mm, clip=true, scale=.8]{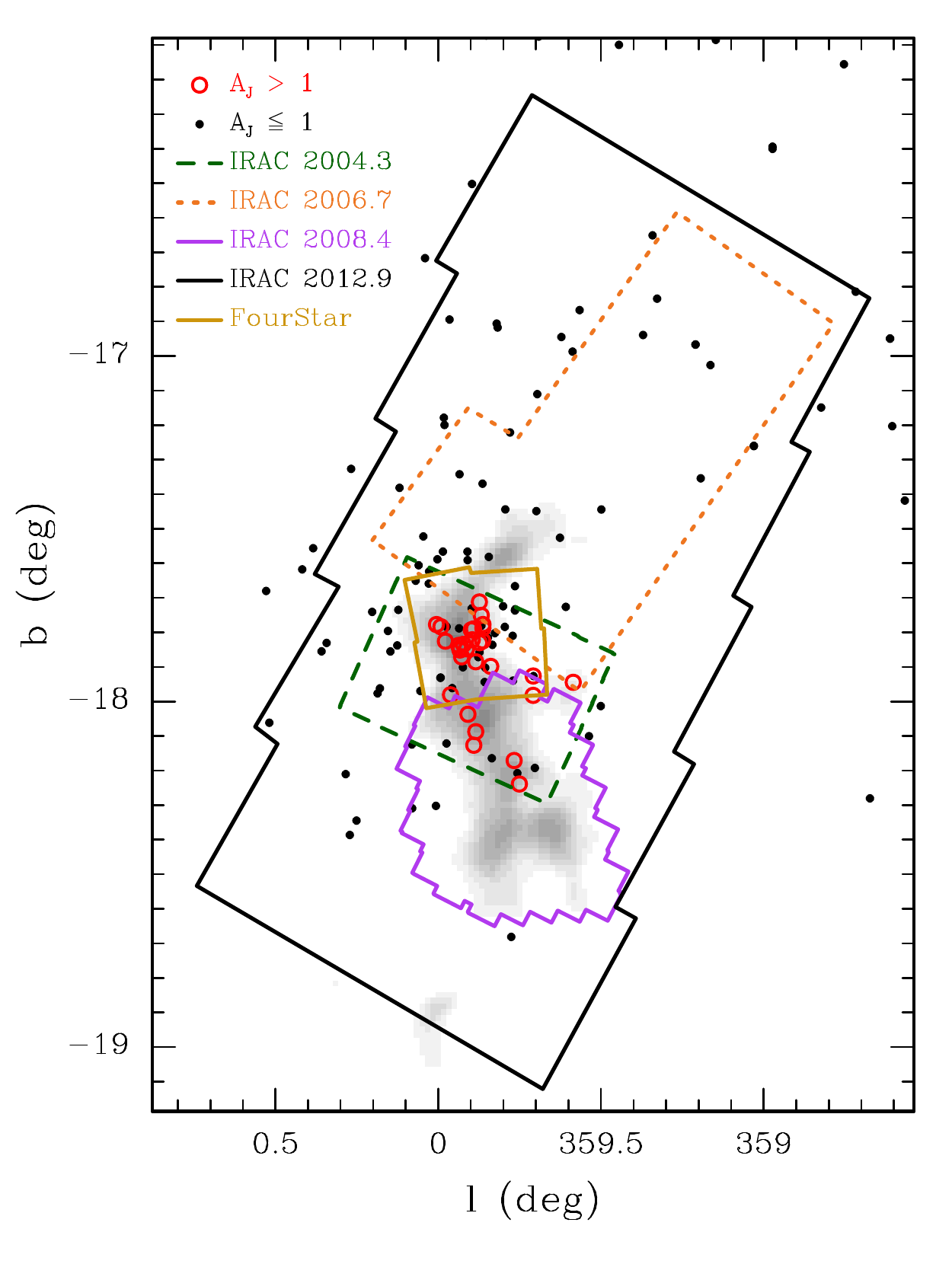}
\caption{
Map of the fields toward Corona Australis that have been imaged by IRAC 
on Spitzer (Table~\ref{tab:epochs}) and FourStar on Magellan.
The known members are plotted with symbols based on their extinction.
The Corona Australis cloud is represented by 
an extinction map \citep[gray scale;][]{dob13}.
}
\label{fig:iraccoverage}
\end{figure}

We began by compiling all [3.6] and [4.5] images
that encompass the Corona Australis molecular cloud.
The Astronomical Observing Requests (AORs), program identifications (PIDs), 
and principle investigators (PIs) of the observations are listed
in Table~\ref{tab:epochs}. These data have primarily been used to classify
the circumstellar disks of previously known members of the cloud
\citep{sic08,cur11,pet11}. 
The IRAC images were obtained at four epochs that span 8.6 years. 
The exposure time for individual images was 10.4~s.
For each of the first three epochs, two exposures were taken in each of the
two bands for a given position. In the fourth epoch, nine images were obtained
in each band. In Figure~\ref{fig:iraccoverage}, we have plotted a map with 
the fields encompassed by the four epochs of IRAC imaging.

We have measured astrometry for the sources in the IRAC images in
[3.6] and [4.5] using the methods developed in \citet{esp16} and \citet{esp17}.
First, the pixel positions, fluxes ($F_\nu$), and signal-to-noise ratios
(S/Ns) for every source were measured using the
point-response-function fitting routine in the Astronomical 
Point source Extractor \citep[APEX;][]{mak05}.
We ignored sources with less than three detections from among all [3.6] and
[4.5] exposures in a given epoch and sources that were close to saturation,
as indicated by $F_\nu$/(exposure time) $>0.73$ and $>0.82$~Jy~s$^{-1}$
in [3.6] and [4.5], respectively.
The pixel positions of the retained sources were corrected for distortion.
Next, we estimated the central world coordinates and orientations 
of the fourth epoch of images using astrometry from Gaia DR2 when available, 
and otherwise used data from the sixth data release (DR6) of the Visible and 
Infrared Survey Telescope for Astronomy (VISTA) Hemisphere Survey 
\citep[VHS,][]{mcm13}. We iteratively calculated a catalog of average 
positions of each source from the individual detections and 
updated the central coordinates and orientations using the new catalog.  
The orientations and central coordinates of the first three epochs were 
measured using the catalog of average positions in the fourth epoch. 

Our proper motion calculations also made use of astrometry from near-IR images 
that we obtained with FourStar on the Magellan I Telescope at Las Campanas 
Observatory \citep{per13}. FourStar contains four arrays with plate scales of 
$0\farcs159$ pixel$^{-1}$ arranged in a square with a total field of
view of $10\farcm8\times10\farcm8$.
We imaged the Coronet Cluster at $J$ and $H$ with four pointings on 
2018 April 3, as shown in Figure~\ref{fig:iraccoverage}. 
Each pointing consisted of 5/18 dithered images with individual exposure times 
of 64/5.8~s for $J$/$H$. 
The images were flat-field corrected and combined, and sources in
the resulting images were identified using routines within IRAF. 
The average FWHM of point sources in the reduced images was $1\farcs1$. 
The completeness limits of the images were $J=20.5$ and $H=19.25$,
which are fainter than the limits from IRAC for the typical colors of 
young brown dwarfs.  For every combined image in a given array, we fit fifth 
order polynomials to correct for distortion using astrometry from Gaia DR2. 

\begin{figure}[h]
  \centering
  \includegraphics[trim = 0mm 0mm 0mm 0mm, clip=true, scale=.6]{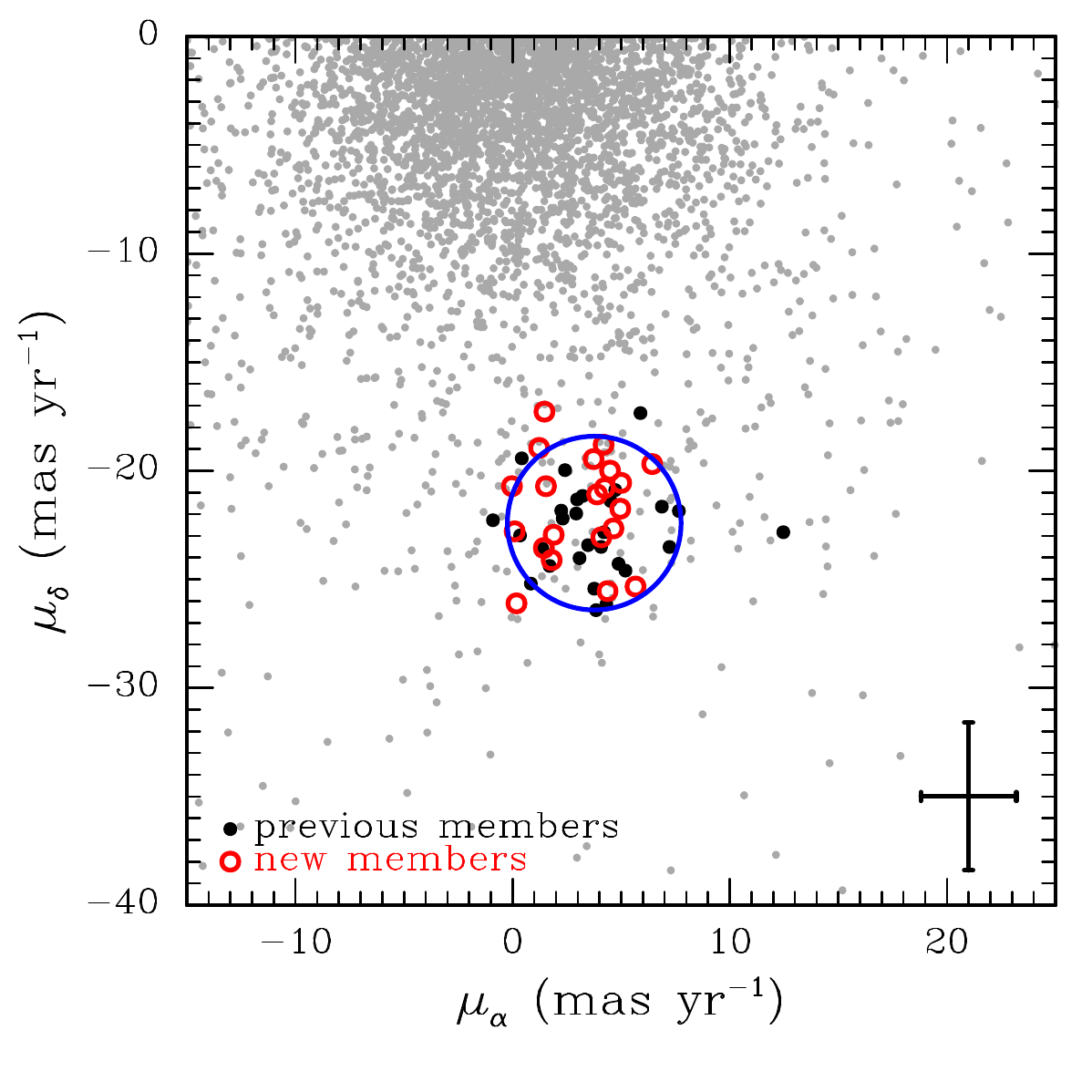}
\caption{
Relative proper motions based on multi-epoch imaging from IRAC
and FourStar for Corona Australis, consisting of previously known members 
(black points), new members from our spectroscopic survey (red circles), 
and all other sources (gray points). The typical errors are indicated.
Sources within 1~$\sigma$ of the blue circle were selected as candidate members.
}
\label{fig:iracpm}
\end{figure}

Relative proper motions were measured from the multiple epochs of astrometry
measured with IRAC and FourStar 
using a linear fit of right ascension and declination as a function of time.  
Because of the longer baseline from the combination of IRAC and FourStar
(available only for the Coronet Cluster) relative to the IRAC epochs alone, 
the proper motion uncertainties from the former are smaller. 
For instance, objects with $J$ = 14/17 have typical errors of  
2.3/8.0~mas~yr$^{-1}$ for IRAC+FourStar and 3.5/16.5~mas~yr$^{-1}$ for IRAC
alone. We restrict our analysis to measurements with $\sigma_\alpha$ and 
$\sigma_\delta<10$~mas~yr$^{-1}$. 
We have measured IR proper motions for 49 of the adopted members of
Corona Australis in Table~\ref{tab:mem}, 28 of which have
previous spectral classifications and 21 of which are newly classified in
this work. Those motions are included in Table~\ref{tab:mem}.
We also have plotted the motions of those members and all other sources
in our proper motion catalog in Figure~\ref{fig:iracpm}.
The members are well-separated from most other sources and have a median 
motion of ($\mu_\alpha$, $\mu_\delta$) = (3.75,$-$22.47)~mas~yr$^{-1}$. 

We have used the IR proper motions to identify candidate members of 
Corona Australis. We selected objects with proper motions within
1~$\sigma$ of a 4~mas~yr$^{-1}$ radius from the median motion of known members, 
which is shown in Figure~\ref{fig:iracpm}. 
This threshold was selected because it is large enough to
recover most of the previously known members.
This criterion produces 241 objects that lack previous spectra, most of
which do not satisfy the photometric criteria in the next section.
All adopted members in Table~\ref{tab:mem} that have IR proper motions
satisfy the criterion with the exception of CrA-26. 
The latter is projected against the molecular cloud and has high enough 
extinction that it is unlikely to be in the foreground of the cloud. 
Its proper motion from Gaia is more accurate than the IR measurements 
and is discrepant by only 1.5~$\sigma$, so we have adopted it as a member.

\subsection{Color-magnitude Diagrams}
\label{sec:cmd}

To refine the kinematic candidates identified with Gaia and 
IR imaging and to select additional candidates that lack kinematic data,
we have utilized several CMDs constructed from optical and IR photometry.
We have made use of photometry from Gaia EDR3 ($G$), 
VHS DR6 ($J$ and $K$), 2MASS ($J$, $H$, $K_s$), AllWISE (W1 and W2),
FourStar ($J$ and $H$), and IRAC ([3.6]).
Offsets were applied to the VHS photometry to align it to the 2MASS data
in the manner described by \citet{luh20}.
Data from Gaia, VHS, and 2MASS are available for the entire survey field.
The areas imaged by FourStar and IRAC are shown in
Figure~\ref{fig:iraccoverage}.
In addition, we obtained $z$-band images of the Coronet Cluster
with the Inamori Magellan Areal Camera and Spectrograph \citep[IMACS,][]{dre11}
on the Magellan I telescope on 2018 April 4.
The instrument was operated in the f/2 mode, 
which produces a plate scale of $0\farcs20$ pixel$^{-1}$
and a circular field of view with a diameter of 15$\arcmin$.
The observations consisted of five dithered 150~s exposures.
The average FWHM for point sources in the IMACS images was $1\arcsec$.

To merge the catalogs of photometry, we identified all matching sources
and adopted the photometry with the smallest errors when multiple similar
bands were available.
In our analysis, we have omitted photometry with $\sigma>0.1$~mag.
To enable the construction of extinction-corrected CMDs,
we estimated the extinction for each source using near-IR photometry
in the manner done in our previous surveys of star-forming regions
\citep[][references therein]{esp17}.
The data in the CMDs were then corrected for extinction using the reddening
relations of \cite{sch16} and \cite{ind05}.

\begin{figure}[h]
  \centering
  \includegraphics[trim = 0mm 0mm 0mm 0mm, clip=true, scale=.6]{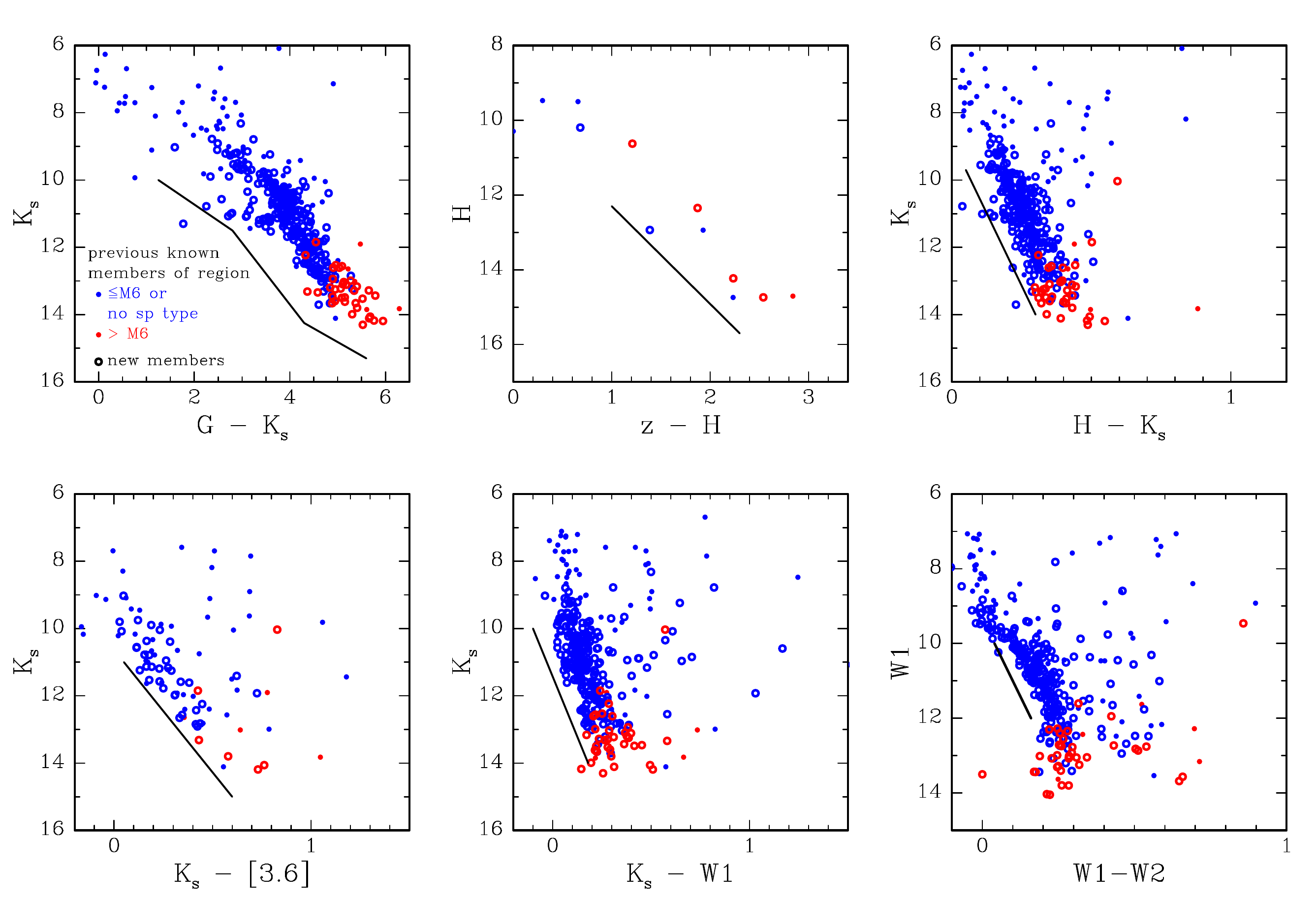}
\caption{
Extinction-corrected CMDs of the previously known members of Corona Australis
(points) and new members from our spectroscopic survey (circles). 
We have selected candidate members based on positions above the solid 
boundaries.
}
\label{fig:criteria}
\end{figure}

In Figure~\ref{fig:criteria}, we show the six extinction-corrected
CMDs that we used to identify candidate members of Corona Australis. 
In each CMD, we have indicated a boundary that follows the lower envelope
of the sequence of known members.
We selected objects as photometric candidates if they appeared above a boundary
in at least one CMD and were not below a boundary in any CMD.
Among the 346 Gaia kinematic candidates from Section~\ref{sec:pp}, 311 satisfy 
the CMD criteria. In Section~\ref{sec:iracpm}, we
identified 241 proper motion candidates from IR imaging that lack
previous spectra, 13 of which are also among the Gaia kinematic candidates.
Among the remaining 228 IR proper motion candidates, 22 satisfy the CMDs.

\section{Spectroscopy of Candidate Members}
\label{sec:spec}

\subsection{Observations}

We obtained optical and/or near-IR spectra of the most promising
candidate members of Corona Australis to measure their spectral types
and assess their youth.
Highest priority was given to candidates that are located
within the FourStar and IRAC fields (regardless of whether kinematic data
were available) and candidates at any location in the full survey field that
have kinematic data from Gaia.  We also observed a small sample of 
stars that have spectral classifications from previous studies.
The spectroscopic sample contains a total of 365 objects, which consist
of 42 stars with previous classifications that are among our adopted members
in Table~\ref{tab:mem}, 293 new members that are spectroscopically
classified as young for the first time in this work, and 30 sources
that are non-members based on either their spectra or kinematics.
Some of the latter were observed before data from
Gaia DR2 and EDR3 became available.
Several objects were observed with both optical and IR spectrographs.

The spectra were collected with the Cerro Tololo Ohio State Multi-Object
Spectrograph (COSMOS), which is based on an instrument described by
\citet{mar11}, and the Astronomy Research using the Cornell Infra-Red
Imaging Spectrograph \citep[ARCoIRIS;][]{sch14} on the 4~m Blanco telescope
at Cerro Tololo Inter-American Observatory (CTIO);
the Folded-port InfraRed Echellette \citep[FIRE;][]{sim13}
on Magellan I at Las Campanas Observatory;
the Gemini Multi-Object Spectrograph \citep[GMOS;][]{hoo04} and
FLAMINGOS-2 on the Gemini South Telescope \citep{eik04}; 
and SpeX \citep{ray03} at the NASA Infrared Telescope Facility (IRTF).
The instrument configurations, dispersers, wavelength 
coverages, and resolutions are summarized in Table~\ref{tab:log}.
The observation dates for the individual targets are listed in 
Table~\ref{tab:spec}.

\begin{figure}[h]
  \centering
  \includegraphics[trim = 0mm 0mm 0mm 0mm, clip=true, scale=1]{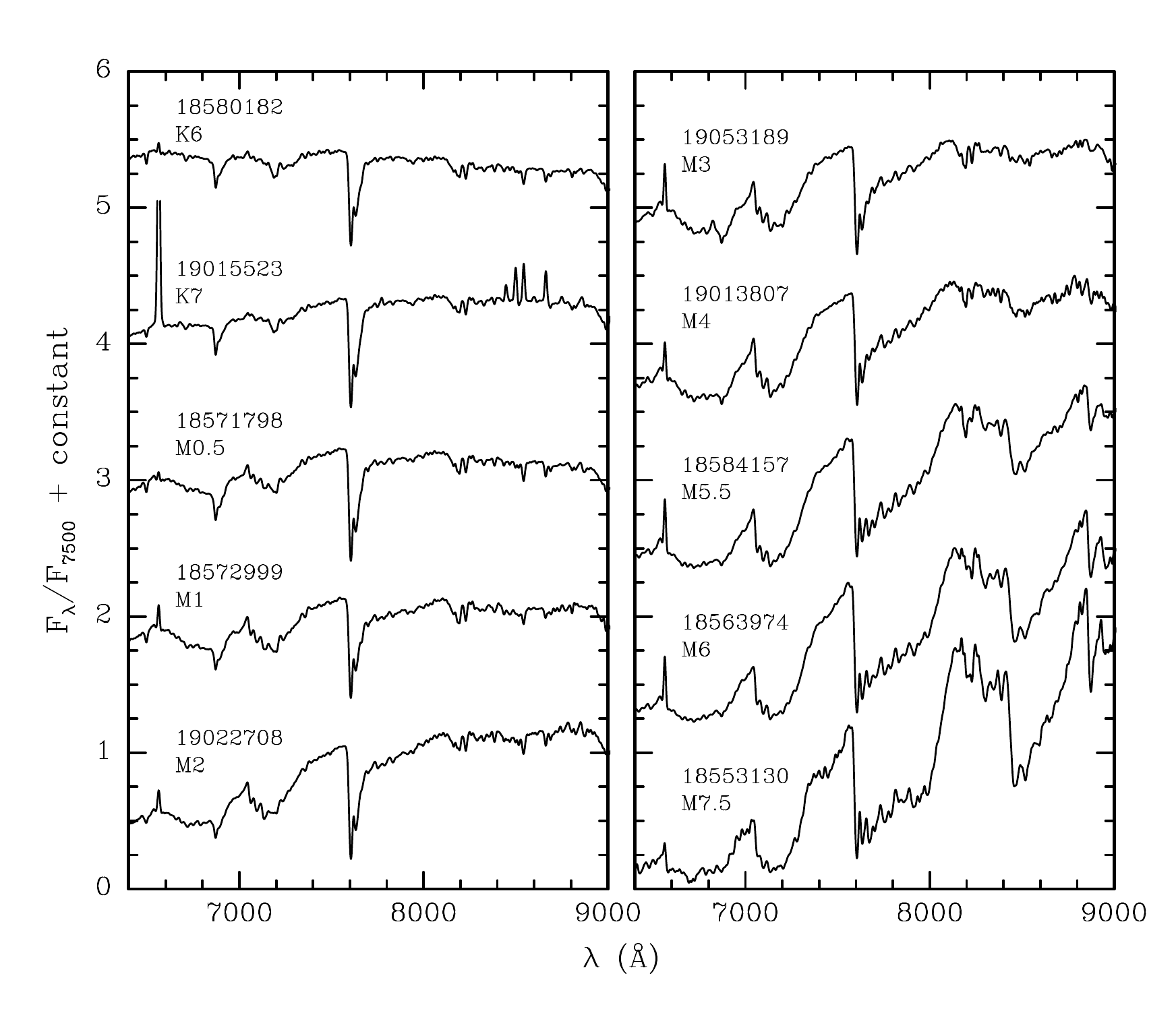}
\caption{
Examples of optical spectra of new members of Corona Australis 
(Table~\ref{tab:spec}). The spectra have been smoothed to a resolution of 
13~\AA. The data used to create this figure are available.
}
\label{fig:optfig}
\end{figure}

\begin{figure}[h]
  \centering
  \includegraphics[trim = 0mm 0mm 0mm 0mm, clip=true, scale=1]{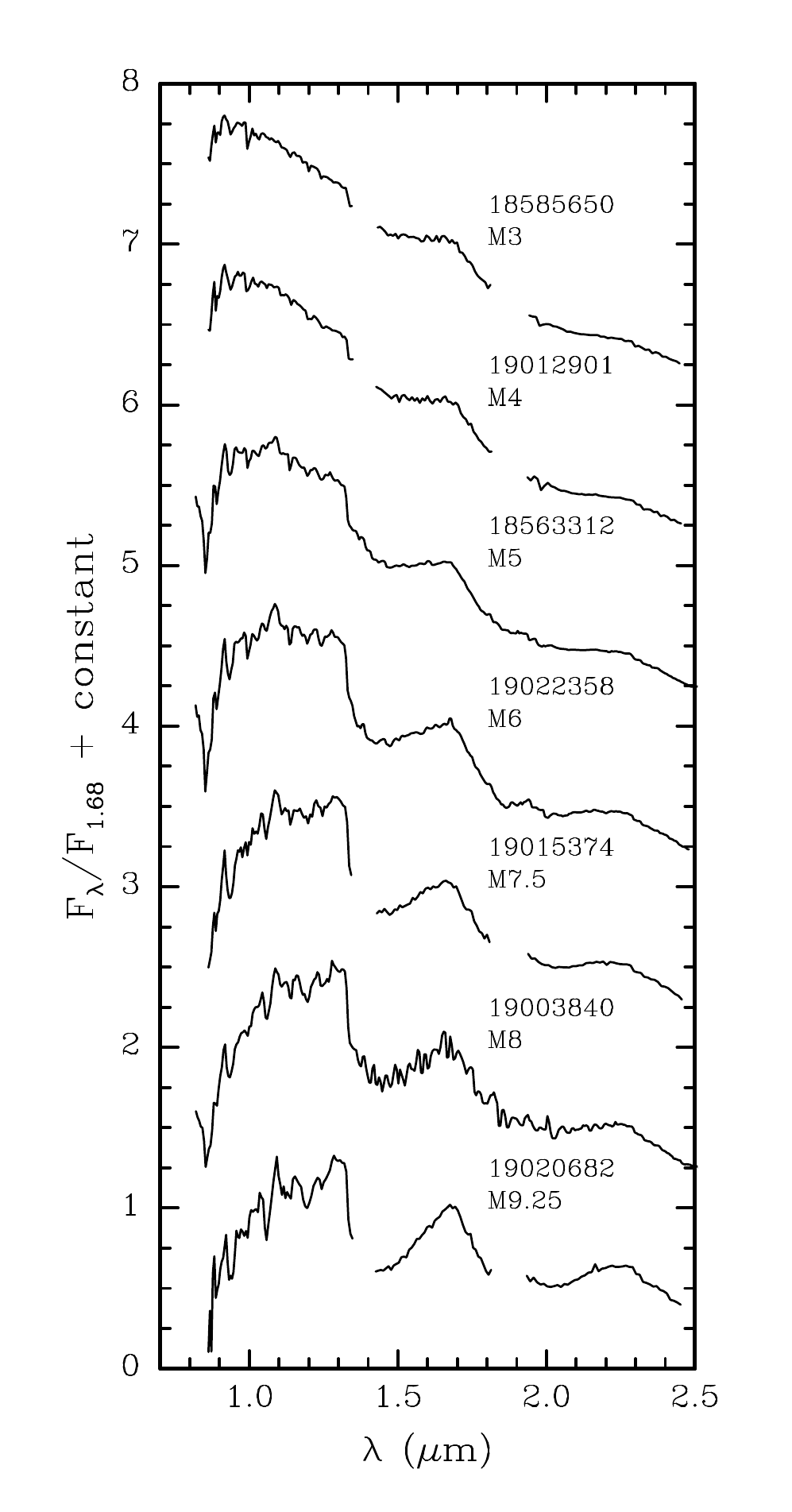}
\caption{
Examples of IR spectra of new members of Corona Australis 
(Table~\ref{tab:spec}). The spectra have been dereddened to match the slopes 
of young standards from \cite{luh17}. 
The data used to create this figure are available.
}
\label{fig:irfig}
\end{figure}

The optical spectra from COSMOS and GMOS was reduced using routines in IRAF.
We applied a flat field correction to the images, extracted the spectra,
and performed the wavelength calibration using arc lamp spectra.
The IR spectra from SpeX and ARCoIRIS were reduced using the Spextool 
package \citep{cus04} and a modified version of Spextool, respectively. 
Spextool corrects for telluric absorption in the manner described by 
\citet{vac03}.
The FIRE and FLAMINGOS-2 data were processed with steps similar to those
in Spextool using routines in R and IRAF, respectively.
We present examples of the optical and IR spectra in Figures~\ref{fig:optfig}
and \ref{fig:irfig}, respectively. The reduced spectra are provided 
in electronic files associated with those figures.

\subsection{Spectral Classification}
\label{sec:class}

The photometry of our candidates indicates that they are likely to have
K/M spectral types if they are members of Corona Australis. 
At those types, diagnostics of youth include Li~I absorption at 6707~\AA\ and
gravity-sensitive features like the Na~I double near 8190~\AA\ and 
the H$_2$O absorption bands at near-IR wavelengths \citep{mar96,luh97,luc01}. 
Spectroscopic evidence of youth is found for 340 of the 365 targets,
which consist of 42 previously classified stars that are among
our adopted members, 293 new members that are classified as young for
the first time with our spectra, and five stars that are rejected
as members based on their kinematics. As noted in the previous section, 
some of the spectroscopic targets classified as non-members would not
satisfy our final selection criteria using Gaia EDR3.
In Table~\ref{tab:spec}, we indicate whether each object is classified
as young and whether it is adopted as a member of Corona Australis.

In addition to assessing youth and membership,
we have used our spectra to measure spectral types.
Objects that lacked evidence of youth were classified through comparison 
to dwarf standards \citep{hen94,kir91,kir97,cus05,ray09}.
For the young stars, the optical spectra were classified with averages
of dwarf and giant spectra \citep{luh97,luh98,luhsig99} and the IR spectra
were classified with standard spectra constructed from optically classified
young objects \citep{luh17}. 
Our spectral classifications are included in the tabulation of
our spectroscopic sample in Table~\ref{tab:spec}.
The catalog of adopted members of Corona Australis in Table~\ref{tab:mem}
contains 39 objects with spectral types later than M6
\citep[$\lesssim0.1$~$M_\odot$,][]{bar98}, 36 of which have been classified
for the first time in this work.

The most promising remaining candidates that lack spectral classifications
consist of ten stars that satisfy our selection criteria
for Gaia kinematics and CMDs (Sections~\ref{sec:edr3} and \ref{sec:cmd})
and have RUWE$<$1.6. They are listed in Table~\ref{tab:remain}.
In addition, we have included in Table~\ref{tab:remain} two candidates
identified with Gaia kinematics and CMDs that have RUWE$>$1.6
and 11 candidate companions to adopted members ($<2\arcsec$) that satisfy
the CMDs but lack Gaia kinematic data.
Based on their photometry, the first sample of 10 candidates 
should have spectral types that range from $\sim$K2 to M8.

\section{Properties of the Corona Australis Stellar Population}
\label{sec:stellarpop}

\subsection{Kinematics}
\label{sec:uvw}

In Table~\ref{tab:mem}, we have included previous measurements of radial
velocities that have errors less than 4~km~s$^{-1}$, which are available for 
82 adopted members of Corona Australis.
We have adopted an error of 0.4~km~s$^{-1}$ for velocities from
\citet{tor06} for which errors were not reported, which is near the 
typical precision estimated in that study.
For the 72 stars with radial velocities and Gaia measurements of proper
motions and parallaxes, we have used the radial velocities,
proper motions, and parallactic distances \citep{bai21} to compute $UVW$
space velocities \citep{joh87}, which are presented in Table~\ref{tab:mem}.
Errors in the space velocities were estimated using the radial velocity
errors and the covariance matrices of errors and correlation coefficients
for the Gaia astrometry \citep{bro97}.
Among the 55 stars with $UVW$ estimates, $\sigma_\pi<0.5$~mas, and RUWE$<$1.6,
the median velocity is $U$, $V$, $W$, = $-3.9$,$-17.4$,$-9.3$~km~s$^{-1}$.

\begin{figure}[h]
  \centering
  \includegraphics[trim = 0mm 0mm 0mm 0mm, clip=true, scale=.6]{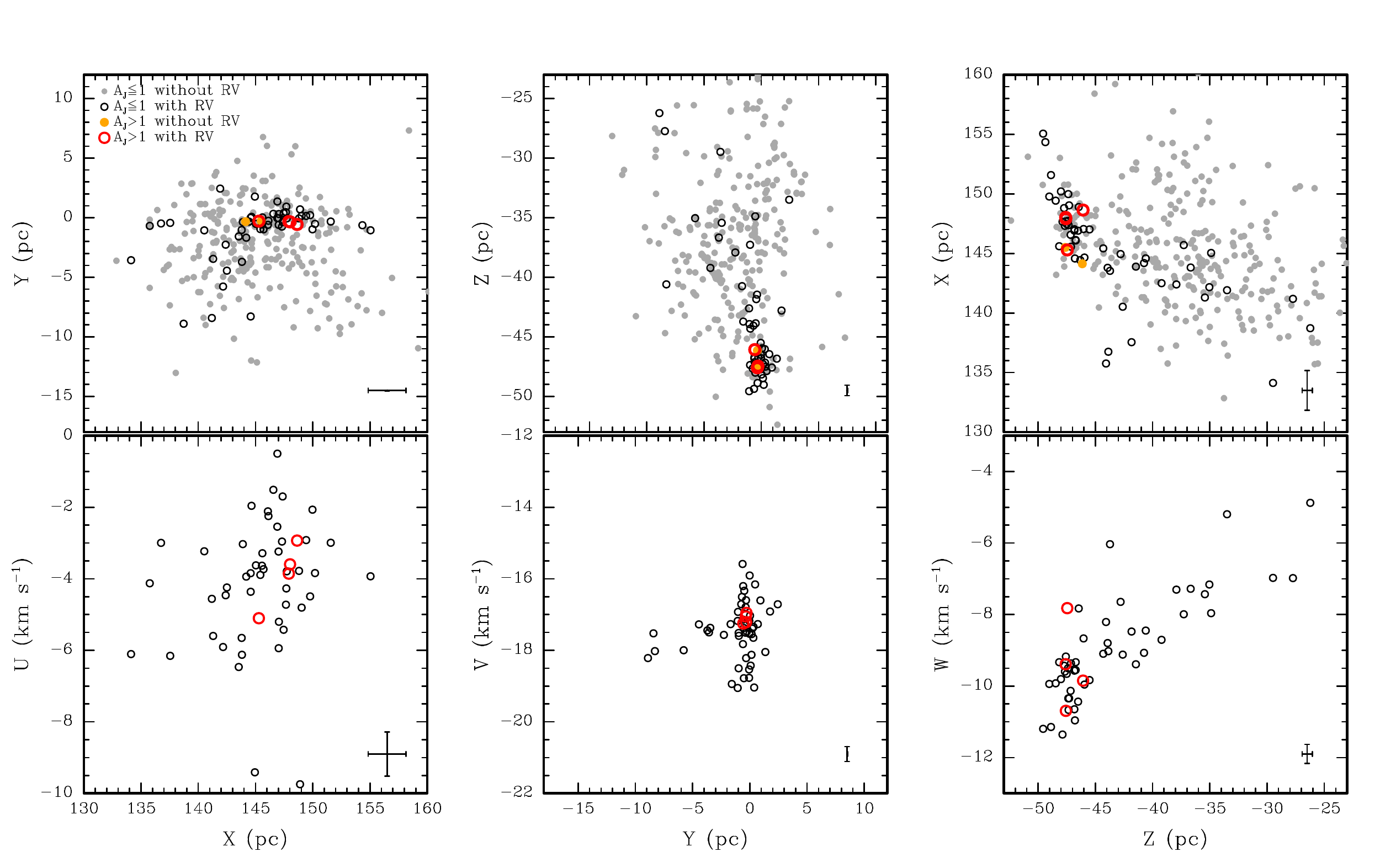}
\caption{
Top: Galactic Cartesian coordinates for members of Corona Australis
with $\sigma_\pi<0.5$ and RUWE$<$1.6 (Table~\ref{tab:mem}).
Bottom: Stars that have radial velocity measurements are plotted in
diagrams of $U$, $V$, and $W$ versus $X$, $Y$, and $Z$, respectively.
The average errors are indicated.
}
\label{fig:uvw}
\end{figure}

\begin{figure}[h]
  \centering
  \includegraphics[trim = 0mm 0mm 0mm 0mm, clip=true, scale=.6]{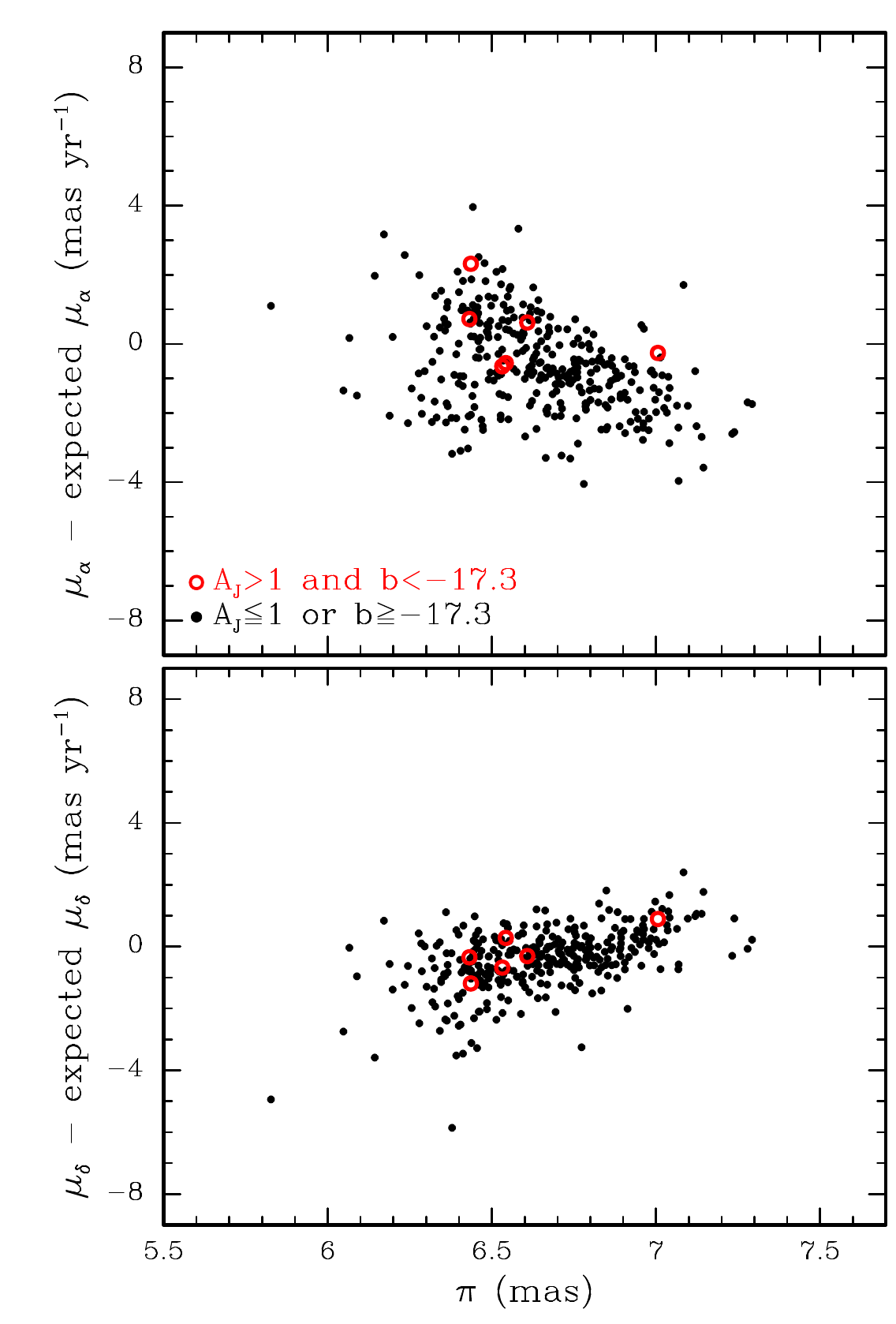}
\caption{
Proper motion offsets versus parallax for members of Corona Australis
with $\sigma_\pi<0.5$ and RUWE$<$1.6 (Table~\ref{tab:mem}).
Different symbols are used for stars embedded in the cloud ($A_J>1$, 
$b<-17.3\arcdeg$) and all other sources ($A_J\leq1$ or $b\geq-17.3\arcdeg$).
}
\label{fig:poppm}
\end{figure}

In the top row of Figure~\ref{fig:uvw}, we have plotted diagrams of 
Galactic Cartesian coordinates ($XYZ$) for adopted members with 
$\sigma_\pi<0.5$~mas and RUWE$<$1.6. Those data indicate that the dimensions of 
Corona Australis range from 15--25~pc. In the bottom row of 
Figure~\ref{fig:uvw},
we show diagrams of $U$, $V$, and $W$ versus $X$, $Y$, and $Z$, respectively, 
for the 55 stars with $UVW$ estimates, $\sigma_\pi<0.5$~mas, and RUWE$<$1.6.
The stars exhibit a correlation between $W$ and $Z$, which indicates expansion
in that dimension, whereas correlations are not evident in the other two
dimensions. In those diagrams of $UVW$ versus $XYZ$, we have plotted stars 
with extinctions above and below $A_J=1$ with different symbols to
compare the kinematics of the stars embedded in the molecular cloud and the 
remaining stars. 
The choice of $A_J=1$ is based on analysis in the next section.
The two samples span similar ranges of $U$, $V$, and $W$. 
To expand the comparison of stars with low and high extinction to include
those lacking radial velocity measurements, we have plotted the proper
motion offsets and parallaxes for adopted members with
$\sigma_\pi<0.5$~mas and RUWE$<$1.6 in Figure~\ref{fig:poppm}.
One symbol is used for stars with $A_J>1$ and $b<-17.3\arcdeg$, which are
likely embedded in the cloud, and a second symbol is used for all other stars.
As in the $UVW$ data, the embedded stars fall within the range of proper 
motion offsets and parallaxes exhibited by the less obscured members.

\subsection{Relative Ages}
\label{sec:age}

As discussed in Section~\ref{sec:intro},
the Coronet Cluster within the Corona Australis cloud appears to have an
age of a few Myr \citep[e.g.,][]{mey09} while evidence has been
reported for the presence of older stars in the vicinity of the cloud.
For example, \cite{caz19} found that a sample of candidate members
exhibited a distribution of dust masses that resembled that of
Upper Sco, which has an age of $\sim$10~Myr \citep{pec12,pec16,luh20}.
Stars older than the Coronet Cluster also have been identified in an area
surrounding the cloud \citep{neu00,pet11}.
Meanwhile, \cite{gal20} found a small difference in ages (5 vs. 6~Myr)
between stars near and far from the cloud among their candidate members 
identified with Gaia DR2 (Section~\ref{sec:kin}).

We can place new constraints on the presence of systematic variations
in age in Corona Australis using our new census.
Low-mass stars are predicted to evolve mostly vertically in the
Hertzsprung-Russell diagram \citep[e.g.,][]{bar98}, so we can estimate the 
relative ages of young stars using their deviations in absolute magnitudes from
a fiducial sequence. 
In previous studies \citep{esp18,esp20}, we have performed analysis of this
kind using diagrams of $M_K$ versus spectral type, estimating 
the difference between the extinction-corrected $M_K$ of a given star and 
the median sequence for Upper Sco, which is denoted by $\Delta M_K$.
We have applied this method to the K5--M5 stars in our census of
Corona Australis.  For stars with $\sigma_\pi<0.5$~mas and RUWE$<$1.6,
we have calculated $M_K$ using the parallactic distances from \citet{bai21}.
For the remaining stars, we have adopted the median distance of members
with $\sigma_\pi<0.5$~mas and RUWE$<$1.6 at $b<-17.3\arcdeg$, which
is where most of the stars lacking accurate parallaxes are located.

\begin{figure}[h]
  \centering
  \includegraphics[trim = 0mm 0mm 0mm 0mm, clip=true, scale=1]{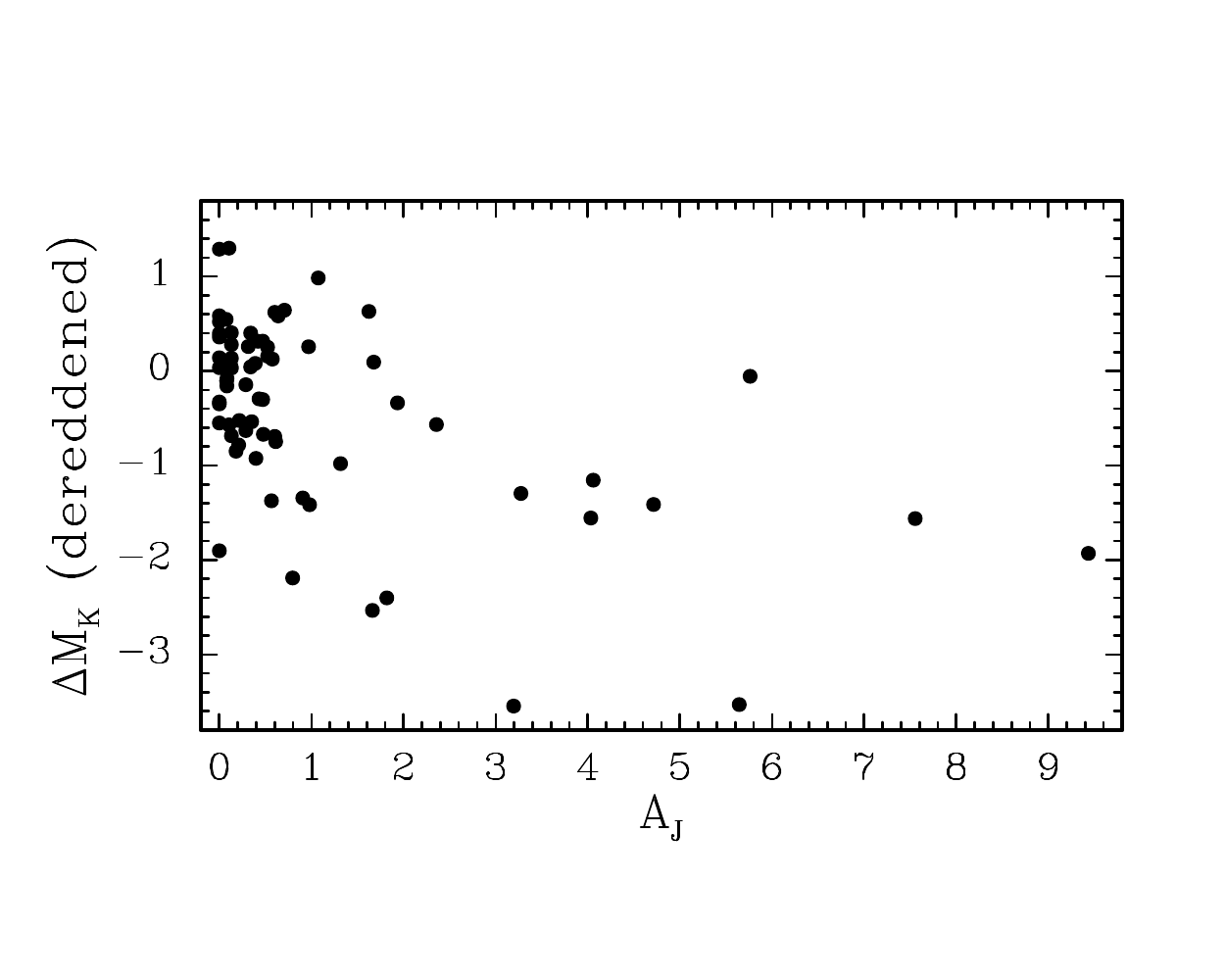}
\caption{
$\Delta M_K$ versus $A_J$ for members of Corona Australis 
that are near the cloud ($b<-17.3$). 
The metric is defined as the difference between the extinction-corrected 
$M_K$ and the median value for Upper Sco members at a given spectral type 
\citep{esp18}. Higher values of $\Delta M_K$ correspond to older ages.
}
\label{fig:ajdeltamk}
\end{figure}

\begin{figure}[h]
  \centering
  \includegraphics[trim = 0mm 0mm 0mm 0mm, clip=true, scale=1]{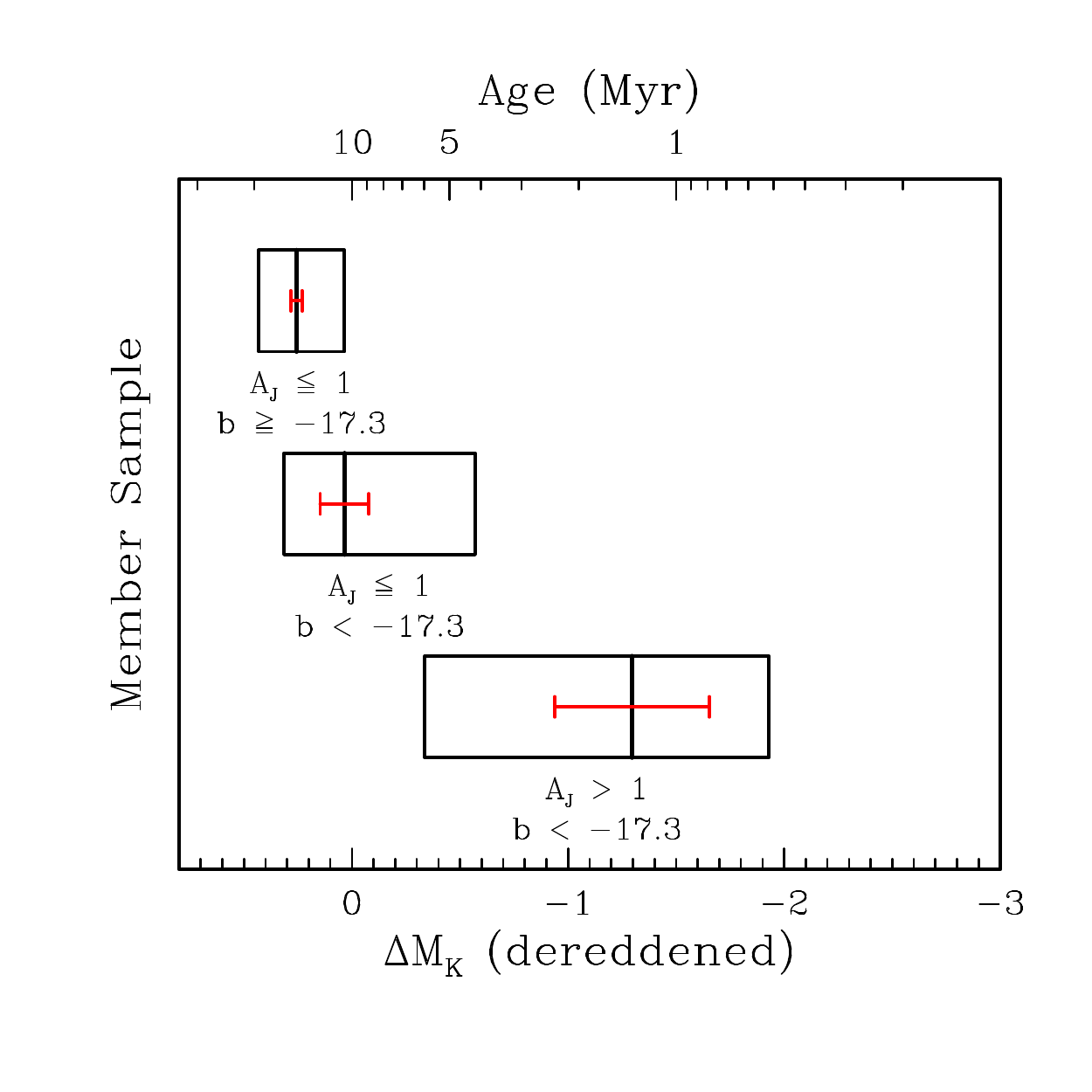}
\caption{
Box diagrams of the interquartile range of $\Delta M_K$ calculated 
for individual members in three samples of Corona Australis members.
The error bars represent the errors on the medians, which have been estimated 
via bootstrapping.
}
\label{fig:box}
\end{figure}

We have examined the variation of $\Delta M_K$ with celestial coordinates 
and extinction. We find that within an area toward
the cloud, stars with extinctions above and below $A_J\sim1$ have 
systematically different $\Delta M_K$, which is illustrated
in Figure~\ref{fig:ajdeltamk}, where we plot $\Delta M_K$
versus $A_J$ for members at $b<-17.3\arcdeg$.
Therefore, we have divided the members of Corona Australis into three
samples for the remaining analysis: stars near the cloud ($b<-17.3\arcdeg$)
with $A_J>1$ (concentrated in the Coronet Cluster), stars near the cloud 
($b<-17.3\arcdeg$) with $A_J\leq1$, and stars farther from the cloud 
($b\geq-17.3\arcdeg$). Nearly all of the latter have $A_J<1$.
In Figure~\ref{fig:box}, we show a box-and-whisker diagram of the
medians and interquartile ranges of $\Delta M_K$ for those three samples. 
Errors in the medians were estimated via bootstrapping. 
We have indicated in Figure~\ref{fig:box} the ages that correspond to
$\Delta M_K$ assuming an age of 10~Myr for Upper Sco and the luminosity
evolution predicted by evolutionary models \citep{bar15,fei16}.
The near-cloud $A_J>1$ stars exhibit a significantly 
younger median age ($\sim1$--2~Myr) than the less obscured near-cloud stars
($\sim12$~Myr) and the off-cloud stars ($\sim15$~Myr).
These results indicate that two populations are projected against the cloud,
a younger one in which many of the members are embedded in the cloud and
an older one surrounding the cloud.
It is unclear whether the latter population is truly younger than 
the off-cloud stars as implied by Figure~\ref{fig:box}.
It is possible that the young population associated with the cloud
includes less-obscured members that contaminate the near-cloud $A_J\leq1$
sample, which could explain why the latter has a slightly younger median
age than the off-cloud sample.  The near-cloud $A_J\leq1$ sample and the 
off-cloud sample are sufficiently similar
in age that we propose that they are both part of a single older population
(approximated by members with $A_J\leq1$) that surrounds and extends beyond
the cloud.
Since some of the off-cloud stars have been previously referred to 
as ``Upper Corona Australis" \citep{gag18,gal20}, 
we adopt that name for the entire older population. 

The Coronet Cluster and Upper Corona Australis resemble Ophiuchus and
Upper Sco in terms of their relative numbers of stars, spatial distributions, 
kinematics, and ages.
In each region, 1) a population with an age of a few Myr is partially
embedded within a molecular cloud and a population with an age of 
$\gtrsim10$~Myr surrounds the cloud and extends across a much larger area;
2) the cloud and its younger population are located near the edge of 
older population;
3) the older population is substantially richer; 
and 4) the older population spans a wider range of kinematics (proper motion
offsets and $UVW$) that largely encompasses the narrower range of kinematics
of the younger population \citep[Figures~\ref{fig:uvw} and
\ref{fig:poppm},][]{luh21}.

\subsection{Initial Mass Function}

\label{sec:imf}
\subsubsection{Completeness}

\begin{figure}[h]
  \centering
  \includegraphics[trim = 0mm 0mm 0mm 0mm, clip=true, scale=.6]{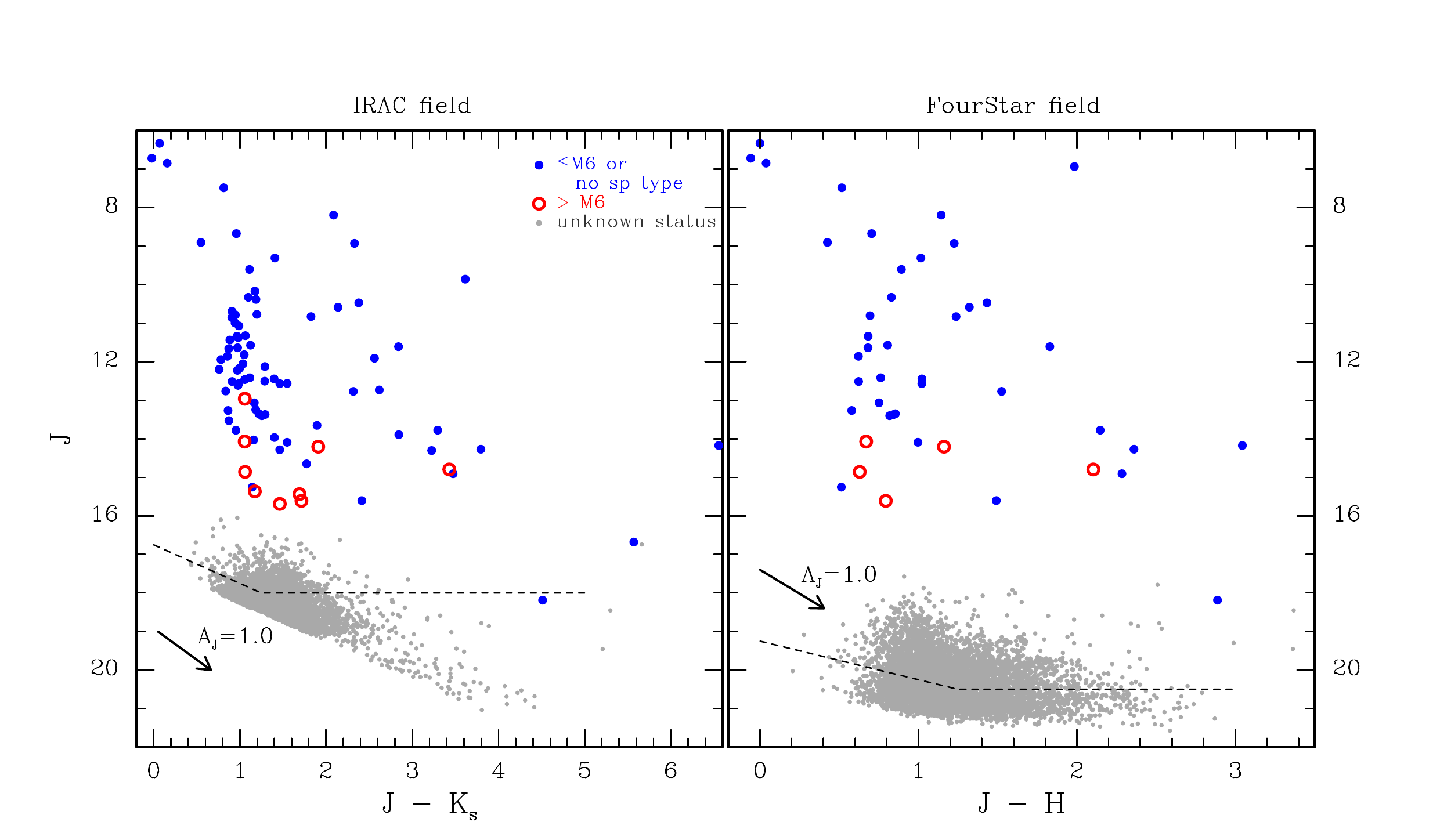}
\caption{
Color-magnitude diagrams of members of Corona Australis with spectral types
$\leq$M6 (blue points) and $>$M6 (red circles) within the field imaged by 
IRAC in multiple epochs (left) and the field imaged by FourStar (right;
see Figure~\ref{fig:iraccoverage}). Sources with unknown status are
indicated (gray points). The completeness limits for these data are marked
(dashed lines).
}
\label{fig:remain}
\end{figure}

To analyze the IMF in Corona Australis, 
we begin by characterizing the completeness of our sample of adopted members
within the field imaged by IRAC in two epochs or more and the field
imaged by FourStar (Figure~\ref{fig:iraccoverage}).
We have constructed a CMD for the IRAC field using $J$ and $K_s$ from 2MASS 
and VHS and a CMD for the FourStar field using $J$ and $H$ from 2MASS and
FourStar. These bands were chosen because they have the 
greatest sensitivity to brown dwarfs for these fields. 
In Figure~\ref{fig:remain}, we show CMDs for the adopted members of
Corona Australis in the IRAC and FourStar fields and the remaining sources 
that are not rejected by spectroscopy or any of the photometric and kinematic 
criteria in Section~\ref{sec:cand}.
For the IRAC and FourStar fields, there are no remaining sources with
undetermined status down to extinction-corrected magnitudes of 
$J=15.6$ and 17, respectively, for $A_J<1$.
Assuming an age of 10~Myr, the median 
parallactic distance of the known members within the field,
and the bolometric corrections from \cite{fil15}, $J=15.6$ and 17 should
correspond to masses of $\sim$0.02 and $\sim$0.015~$M_\odot$, respectively,
according to evolutionary models \citep{bur97,cha00}.
For extinctions high enough to reach most members of the Coronet Cluster
($A_J<5$), the completeness limit is near an extinction-corrected magnitude
of $J=14$ in the FourStar field, which corresponds to 0.03--0.04~$M_\odot$ 
for ages of 1--3~Myr.
When using Gaia EDR3 to search for members of Upper Sco, 
\citet{luh21} found that the completeness was high among spectral types 
earlier than $\sim$M7 for the low extinctions in the association ($A_J<1$). 
If we account for the differences in age and distance 
between Upper Sco and Corona Australis, then Gaia EDR3 should provide
a completeness limit of $\sim$M6 ($\sim0.1$~$M_\odot$) 
for our full survey field. 
Based on their photometry, roughly half of the 10 
kinematic candidates with RUWE$<$1.6 that lack spectra (Table~\ref{tab:remain})
are expected to have types of $\lesssim$M6.

\subsubsection{Distribution of Spectral Types}

\begin{figure}[h]
  \centering
  \includegraphics[trim = 0mm 0mm 0mm 0mm, clip=true, scale=.8]{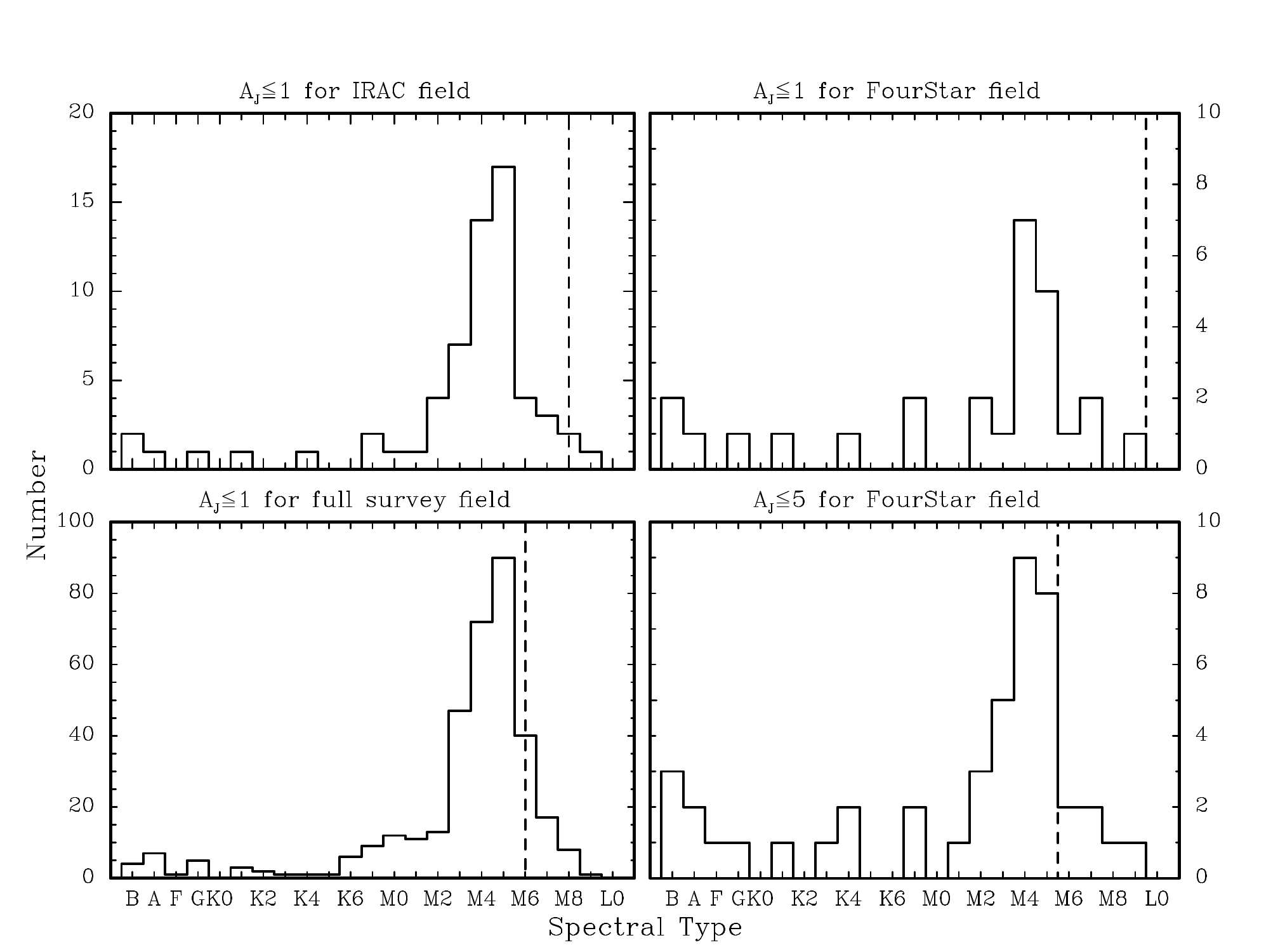}
\caption{
Histograms of spectral types for four samples of adopted members of 
Corona Australis.
The completeness limits are indicated (dashed lines). 
}
\label{fig:imf}
\end{figure}

As done in previous studies of nearby star-forming regions 
\citep[e.g.,][]{esp19}, we have used spectral type as a proxy for stellar 
mass when characterizing the IMF of Corona Australis.
In Figure~\ref{fig:imf}, we show histograms of spectral type 
for the adopted members within the following four samples:
1) $A_J\leq1$ and within the field imaged by IRAC in multiple epochs,
2) $A_J\leq1$ and within our full survey field,
3) $A_J\leq1$ and within the FourStar field; 
and 4) $A_J\leq5$ and within the FourStar field.
In each histogram, we have marked the spectral type that corresponds to
the completeness limit estimated in the previous section for that sample.
The first two samples should be dominated by the members of
the older population, Upper Corona Australis, due to the shallow extinction
limits and the moderate-to-large sizes of the fields.
For the third sample, the small field size favors members of the Coronet
Cluster while the low extinction limit favors members of Upper Corona Australis.
Since the fourth sample applies to the same small field but extends to
higher extinction, it likely consists primarily of Coronet members.
All of the histograms in Figure~\ref{fig:imf} exhibit a peak near M5
($\sim0.15$~$M_\odot$), which is consistent with the distributions that
we have measured in other star-forming regions \citep{luh16,esp19,luh21}.

\subsection{Circumstellar Disks}
\label{sec:disks}

\subsubsection{Mid-IR Photometry}

To search for evidence of circumstellar disks among our adopted
members of Corona Australis, we have compiled their mid-IR 
photometry from WISE and Spitzer.
The WISE data were taken in bands centered at 3.4, 4.6, 12, and 22~$\mu$m 
(W1--W4) and the Spitzer data were collected in the four bands of IRAC
([3.6]--[8.0]) and the 24~$\mu$m band ([24]) of the Multiband Imaging 
Photometer for Spitzer \citep[MIPS;][]{rie04}.
For each member, we searched for counterparts in the
AllWISE Source Catalog and the WISE All-Sky Source Catalog.
We adopted the data from the former when available (368 stars) and otherwise 
used the data from the latter (three stars). The WISE photometry is included
in Table~\ref{tab:mem}.
For each member with a WISE counterpart, we visually
inspected the WISE Atlas images to check for false or unreliable detections,
blending with nearby sources, and contamination from extended emission,
which are flagged in Table~\ref{tab:mem}.
Measurements with W2$<6$ were excluded from our analysis 
because they are subject to large systematic errors \citep{cut12}. 
Among the members with WISE counterparts, 98, 97, 84, and 24\% have 
photometry at W1--W4, respectively.

A subset of the members of Corona Australis are encompassed by images
obtained by IRAC and MIPS on Spitzer (Figure~\ref{fig:iraccoverage}).
When available, we retrieved photometry for members from the Spitzer 
Enhanced Imaging Products (SEIP) Source List.
For members that were absent from SEIP but that were detected by
Spitzer based on inspection of the images, we measured photometry
from the SEIP mosaics in the manner described by \citet{esp18}.
In Table~\ref{tab:mem}, we present photometry for 109, 111, 92, 81, and 87
sources in [3.6], [4.5], [5.8], [8.0], and [24], respectively.
All members encompassed by the IRAC images were detected in at least one band 
except for S CrA B, which was unresolved from its companion.
Nine members fall within the MIPS images but were undetected or unresolved
(including S CrA B).

\subsubsection{Disk Detections and Classifications}

As done in our previous surveys \citep[e.g.,][]{esp18}, we have used
extinction-corrected colors between $K_s$ and W2, W3, W4, [4.5], [8.0],
and [24] to detect IR excesses from disks.
For each color, we have subtracted the typical intrinsic value for a young
stellar photosphere at the spectral type of a given star \citep{luh21}.
If a star appears to exhibit an excess in a given band but
a detection in any band at a longer wavelength is consistent with
a photosphere, an excess is not assigned to the first band.
In Table~\ref{tab:mem}, we have included flags to indicate whether
excesses are present in each of the six bands that we have considered.
Excesses are detected for 122 adopted members of Corona Australis.
The excesses for 41 stars have not been identified previously 
\citep{sic08,cur11,pet11,gal20}.

We have used the sizes of the IR excesses to classify the evolutionary stages
of the disks according to the criteria that we have applied in previous studies
\citep{luhM12,esp14,esp18}.
Those stages consist of the following \citep{ken05,rie05,her07,luh10,espa12}:
{\it full disks} are optically thick at mid-IR wavelengths
with no large gaps or holes; {\it transitional disks} are optically thick 
with a large inner hole; {\it evolved disks} are optically thin and
lack large gaps or holes; {\it evolved transitional disks}
are optically thin with a large inner hole; and {\it debris disks} consist 
of second-generation dust produced by collisions of planetesimals.
Debris and evolved transitional disks are indistinguishable in mid-IR
photometry. In Table~\ref{tab:mem}, we have listed the disk classifications 
for the 122 stars that exhibit IR excesses.
We classify 76 as full \citep[including nine previously identified 
protostars,][]{nut05,for06}, 15 as evolved, 28 as debris or evolved 
transitional, two as transitional, and one as evolved or transitional.
In addition, we have appended ``I" to the disk classifications in
Table~\ref{tab:mem} for stars that have been previously classified as 
class~I protostars.

\subsubsection{Disk Fractions}

We have calculated the fraction of adopted members that have full, evolved, 
and transitional disks in each of the three samples defined for the age 
analysis in Section~\ref{sec:age} for four ranges of spectral types that 
correspond roughly to logarithmic intervals of stellar mass \citep{bar98}.
As done in our previous measurements of disk fractions, we have excluded
protostars and evolved transitional disks.
The resulting disk fractions are presented in Table~\ref{tab:diskfraction}.
Several of the disk fractions have large uncertainties due to the small
samples, but we can meaningfully compare the fractions for K6--M5.75, which
offer the largest samples. For that range of types,
the off-cloud, near-cloud $A_J\leq1$, and near-cloud $A_J>1$ samples
have disk fractions of $0.14\pm0.03$, $0.35^{+0.10}_{-0.08}$, 
and $0.61^{+0.25}_{-0.18}$, respectively.
The relative values of those fractions are consistent with the relative ages 
derived for those samples (i.e., the older samples have lower disk fractions).
Similarly, the disk fraction of $0.14\pm0.03$ for the off-cloud sample 
is between the K6--M5.75 disk fractions for Upper Sco ($0.21^{+0.02}_{-0.01}$)
and UCL/LCC \citep[$0.076\pm0.005$][]{luh21b}, which is consistent
with our age estimate for the off-cloud stars ($\sim15$~Myr) relative to
the ages of Upper Sco and UCL/LCC \citep[$\sim10$ and 20~Myr,][]{luh20}.

\section{Conclusion}

We have performed a census of the young stellar populations near
the Corona Australis molecular cloud using photometric and kinematic data
from several sources, particularly Gaia EDR3, and spectroscopy of hundreds
of candidate members. We have used our new catalog of young stars to 
characterize the space velocities, relative ages, IMFs, and circumstellar 
disks in those populations. Our results are summarized as follows:

\begin{enumerate}

\item
For our survey, we have considered a large area that extends well beyond
the Corona Australis cloud ($\alpha=275$ to 288.5$\arcdeg$,
$\delta=-40$ to $-32.5\arcdeg$).
To characterize the kinematics of the young stellar populations
within that field, we have identified a sample of young low-mass stars based 
on their positions in Gaia CMDs and we have applied
a Gaussian mixture model to their proper motion offsets and parallaxes
to separate the populations associated with the cloud from field stars.
The kinematics of the model components for the former were then used as
criteria for selecting candidate members from Gaia EDR3 at all magnitudes
and colors.
In addition, we have identified candidates based on proper motions measured
from multi-epoch IR imaging from the Spitzer Space Telescope and Magellan 
Observatory and positions in CMDs measured with Gaia, 2MASS, VISTA VHS, WISE,
Spitzer, and Magellan.

\item
We have obtained optical and IR spectra of 365 candidate members of
the populations near the Corona Australis cloud, which have been used
to measure spectral types and diagnostics of youth. 
We also have compiled all objects in our survey field
that have evidence of youth and that are not rejected as non-members by
available measurements of proper motions and parallaxes.
The resulting catalog contains 393 adopted members (39 at $>$M6), 293 (36)
of which are spectroscopically classified for the first time in this work.

\item
Previous measurements of radial velocities are available for 82 of the
adopted members. For the 72 stars that also have proper motions and parallaxes
from Gaia EDR3, we have calculated $UVW$ space velocities. 
Among the 55 stars with measured velocities and accurate and reliable
astrometry ($\sigma_\pi<0.5$~mas, RUWE$<$1.6), 
$W$ is correlated with the spatial position in $Z$, which suggests the
presence of expansion along that dimension.
Only a small number of the highly reddened stars ($A_J>1$) within the
Corona Australis cloud have measurements of $UVW$, but those measurements
fall within the range of velocities exhibited by the less obscured members.
To expand the comparison of stars with low and high extinction to include
those lacking radial velocity measurements, we have also considered their
proper motion offsets and parallaxes. In those parameters, as with $UVW$,
the embedded stars fall within the range of values for the less obscured 
members.

\item
We have estimated the relative ages of low-mass stars (K5--M5)
in Corona Australis using their offsets in $M_K$ from the median
sequence for Upper Sco in the H-R diagram.
We calculated the medians of those offsets in three samples of members:
stars near the cloud ($b<-17.3\arcdeg$) with $A_J>1$ (concentrated in
the Coronet Cluster), stars near the cloud ($b<-17.3\arcdeg$) with 
$A_J\leq1$, and stars farther from the cloud ($b\geq-17.3\arcdeg$).
Those median offsets have been converted to ages 
assuming an age of 10~Myr for Upper Sco and the luminosity
evolution predicted by evolutionary models \citep{bar15,fei16}.
We find that the near-cloud $A_J>1$ stars exhibit a significantly
younger median age ($\sim1$--2~Myr) than the less obscured near-cloud stars
($\sim12$~Myr) and the off-cloud stars ($\sim15$~Myr).
The second sample may appear slightly younger than the third one
because of contamination from members of the younger population that have
$A_J<1$. Therefore, we propose that Corona Australis contains two populations,
a younger one that is partially embedded in the cloud (the Coronet Cluster) 
and an older one that surrounds and extends beyond the cloud
\citep[Upper Corona Australis,][]{gag18}.

\item
The Coronet Cluster and Upper Corona Australis resemble Ophiuchus and
Upper Sco in terms of their relative numbers of stars, spatial distributions, 
kinematics, and ages.
In each region, 1) a population with an age of a few Myr is partially
embedded within a molecular cloud and a population with an age of 
$\gtrsim10$~Myr surrounds the cloud and extends across a much larger area;
2) the cloud and its younger population are located near the edge of the
older population; 3) the older population is substantially richer; 
and 4) the older population spans a wider range of kinematics 
that largely encompasses the narrower range of kinematics
of the younger population.

\item
\citet{gal20} identified candidate members of Corona Australis using data
from Gaia DR2. Among those stars, they reported the presence of two
populations -- near the cloud and far from the cloud -- that are distinct
from each other in terms of proper motions. However, we find that the
difference in proper motions is a reflection of the projection effects
that arise for stars distributed across an extended area of sky.
Instead, as mentioned previously, we find that the near-cloud and 
off-cloud stars have similar kinematics in terms of space velocities
and proper motion offsets (which correct for projection effects).

\item
We have used IR CMDs to characterize the completeness limits of our census
of Corona Australis for the areas imaged by Spitzer and Magellan
and the full survey field, arriving at limits that range from
$\sim$0.015--0.1~$M_\odot$ for the levels of extinction encompassing most
known members.
We have constructed histograms of spectral type for extinction-limited
samples of members within multiple fields, each of which exhibits a peak
at M5 ($\sim0.15$~$M_\odot$), indicating that the IMF in Corona Australis
has a similar characteristic mass as other nearby star-forming regions.

\item
We have compiled mid-IR photometry from WISE and Spitzer for the adopted
members of Corona Australis and we have used those data to search for IR
excesses from circumstellar disks.  Excesses are detected for 122 stars, 
a third of which are reported for the first time in this work.
The sizes of the excesses have been used to classify the evolutionary
stages of the disks. The relative disks fractions among the three samples
defined for the age analysis are consistent with the relative ages derived for 
those samples. The same is true for the disk fractions and ages of Upper Corona
Australis ($\sim15$~Myr) relative to those in Upper Sco and UCL/LCC 
($\sim10$ and 20~Myr).

\end{enumerate}

\acknowledgements
K.L. acknowledges support from NASA grant 80NSSC18K0444 for portions of
this work.
We thank Katelyn Allers for providing the modified version of Spextool for
use with ARCoIRIS data. 
The IRTF is operated by the University of Hawaii under contract 80HQTR19D0030
with NASA.
The data at CTIO were obtained through programs 2016A-0157 and 2021A-0006 
at NOIRLab.
CTIO and NOIRLab are operated by the Association of Universities for Research in
Astronomy under a cooperative agreement with the NSF.
The Gemini data were obtained through programs 
GS-2008B-Q-12 (2008B-0185) and GS-2016A-Q-32 (2016A-0139).
Gemini Observatory is operated by AURA under a cooperative agreement with
the NSF on behalf of the Gemini partnership: the NSF (United States), the NRC
(Canada), CONICYT (Chile), Minist\'{e}rio da Ci\^{e}ncia,
Tecnologia e Inova\c{c}\~{a}o (Brazil), Ministerio de Ciencia,
Tecnolog\'{i}a e Innovaci\'{o}n Productiva (Argentina), and
Korea Astronomy and Space Science Institute (Republic of Korea).
This work used data from the European Space Agency (ESA)
mission Gaia (\url{https://www.cosmos.esa.int/gaia}), processed by
the Gaia Data Processing and Analysis Consortium (DPAC,
\url{https://www.cosmos.esa.int/web/gaia/dpac/consortium}). Funding
for the DPAC has been provided by national institutions, in particular
the institutions participating in the Gaia Multilateral Agreement.
This work used data from the Spitzer Space Telescope and the
NASA/IPAC Infrared Science Archive, operated by JPL under contract
with NASA, and the VizieR catalog access tool and the SIMBAD database, both
operated at CDS, Strasbourg, France.
WISE is a joint project of the University of California, Los Angeles,
and the JPL/Caltech, funded by NASA.
2MASS is a joint project of the University of
Massachusetts and the Infrared Processing and Analysis Center (IPAC) at
Caltech, funded by NASA and the NSF.
The Center for Exoplanets and Habitable Worlds is supported by the
Pennsylvania State University, the Eberly College of Science, and the
Pennsylvania Space Grant Consortium.

{\it Facilities: }
\facility{Blanco (COSMOS, ARCoIRIS)},
\facility{Gemini:South (Flamingos-2)},
\facility{IRTF (SpeX)},
\facility{Spitzer (IRAC, MIPS)},
\facility{WISE, Gaia},
\facility{Magellan:Baade (FourStar, FIRE)}.

\bibliographystyle{aasjournal}
\bibliography{refs}

\clearpage

\LongTables
\begin{deluxetable}{ll}
\tabletypesize{\scriptsize}
\tablewidth{300pt}
\tablecaption{Adopted Members of Corona Australis\label{tab:mem}}
\tablehead{
\colhead{Column Label} &
\colhead{Description}}
\startdata
2MASS     &  2MASS Point Source Catalog source name \\
WISE      &  WISE source name\tablenotemark{a} \\
Gaia      &  Gaia EDR3 source name\\
GP75 R CrA & Designation from \cite{gla75} \\
TS84    &  Designation from \cite{tay84} \\
HBC       &  Designation from \cite{her88} \\
WMB97   &  Designation from \cite{wil97} \\
CrAPMS    &  Designation from \cite{wal97} \\
P98c    &  Designation from \cite{pat98} \\
ISO-CrA   &  Designation from \cite{olo99} \\
RX        &  Designation from \cite{neu00} \\
LEM2005b CrA & Designation from \cite{lop05} \\
FP2007  &  Designation from \cite{for07} \\
SHJ2008 &  Designation from \cite{sic08} \\
GMM2009 &  Designation from \cite{gut09} \\
PCB2011 CrA & Designation from \cite{pet11} \\
OName & Other common identifier\\
RAdeg & Right Ascension (ICRS) \\
DEdeg & Declination (ICRS) \\
Ref-Pos & Reference for RAdeg and DEdeg\tablenotemark{b} \\
SpType & Spectral type \\
r\_SpType & Reference for SpType\tablenotemark{c} \\
Adopt & Adopted spectral type \\
GaiapmRA & Proper motion in right ascension from Gaia EDR3\\
e\_GaiapmRA & Error in GaiapmRA \\
GaiapmDec & Proper motion in declination from Gaia EDR3\\
e\_GaiapmDec & Error in GaiapmDec \\
plx & Parallax from Gaia EDR3\\
e\_plx & Error in plx \\
RUWE & Re-normalized unit weight error from Gaia EDR3 \\
kin & Satisfies Gaia kinematic criteria? \\
r\_med\_geo & Median of the geometric distance posterior from \citet{bai21}\\
r\_lo\_geo & 16th percentile of the geometric distance posterior from \citet{bai21}\\
r\_hi\_geo & 84th percentile of the geometric distance posterior from \citet{bai21}\\
IRpmRA & Proper motion in right ascension from IR images \\
e\_IRpmRA & Error in IRpmRA \\
IRpmDec & Proper motion in declination from IR images \\
e\_IRpmDec & Error in IRpmDec \\
RVel & Radial velocity \\
e\_RVel & Error in RVel \\
r\_RVel & Radial velocity reference\tablenotemark{d} \\
U & $U$ component of space velocity \\
e\_U & Error in U \\
V & $V$ component of space velocity \\
e\_V & Error in V \\
W & $W$ component of space velocity \\
e\_W & Error in W \\
Gmag & $G$ magnitude from Gaia EDR3\\
e\_Gmag & Error in Gmag \\
GBPmag & $G_{\rm BP}$ magnitude from Gaia EDR3\\
e\_GBPmag & Error in GBPmag \\
GRPmag & $G_{\rm RP}$ magnitude from Gaia EDR3\\
e\_GRPmag & Error in GRPmag \\
Jmag & $J$ magnitude \\
e\_Jmag & Error in Jmag \\
r\_Jmag & Reference for Jmag\tablenotemark{e}\\
Hmag & $H$ magnitude \\
e\_Hmag & Error in Hmag \\
r\_Hmag & Reference for Hmag\tablenotemark{e}\\
Kmag & $K_s$ or $K$ magnitude \\
e\_Kmag & Error in Kmag \\
r\_Kmag & Reference for Kmag\tablenotemark{e}\\
3.6mag & Spitzer [3.6] magnitude \\
e\_3.6mag & Error in 3.6mag \\
f\_3.6mag & Flag on 3.6mag\tablenotemark{f} \\
4.5mag & Spitzer [4.5] magnitude \\
e\_4.5mag & Error in [4.5] magnitude \\
f\_4.5mag & Flag on 4.5mag\tablenotemark{f} \\
5.8mag & Spitzer [5.8] magnitude \\
e\_5.8mag & Error in 5.8mag \\
f\_5.8mag & Flag on 5.8mag\tablenotemark{f} \\
8.0mag & Spitzer [8.0] magnitude \\
e\_8.0mag & Error in 8.0mag \\
f\_8.0mag & Flag on 8.0mag\tablenotemark{f} \\
24mag & Spitzer [24] magnitude \\
e\_24mag & Error in 24mag \\
f\_24mag & Flag on 24mag\tablenotemark{f} \\
W1mag & WISE W1 magnitude \\
e\_W1mag & Error in W1mag \\
f\_W1mag & Flag on W1mag\tablenotemark{f} \\
W2mag & WISE W2 magnitude \\
e\_W2mag & Error in W2mag \\
f\_W2mag & Flag on W2mag\tablenotemark{f} \\
W3mag & WISE W3 magnitude \\
e\_W3mag & Error in W3mag \\
f\_W3mag & Flag on W3mag\tablenotemark{f} \\
W4mag & WISE W4 magnitude \\
e\_W4mag & Error in W4mag \\
f\_W4mag & Flag on W4mag\tablenotemark{f} \\
Exc4.5 & Excess present in [4.5]? \\
Exc8.0 & Excess present in [8.0]? \\
Exc24 & Excess present in [24]? \\
ExcW2 & Excess present in W2? \\
ExcW3 & Excess present in W3? \\
ExcW4 & Excess present in W4? \\
DiskType & Disk Type \\
Ak & Extinction in K \\
f\_Ak & Method for estimating Ak\tablenotemark{g}
\enddata
\tablecomments{This table is available in its entirety in a machine-readable form.}
\tablenotetext{a}{Coordinate-based identifications from the AllWISE Source
          Catalog when available. Otherwise, identifications are from
          the AllWISE Reject Table or the WISE All-Sky Catalog.}
\tablenotetext{b}{
  2MASS = 2MASS Point Source Catalog;
  Gaia = Gaia EDR3 (Epoch 2016.0);
  IRAC = derived from IRAC images in this work.
}
\tablenotetext{c}{
1 = This Work;
2 = \cite{tor06};
3 = \cite{hou82};
4 = \cite{neu00};
5 = \cite{caz19};
6 = \cite{wal97};
7 = \cite{pat98};
8 = \cite{bou04};
9 = \cite{her88};
10 = \cite{rom12};
11 = \cite{sic11};
12 = \cite{lop05};
13 = \cite{sic08};
14 = \cite{car07};
15 = \cite{pra03};
16 = \cite{mey09};
17 = \cite{rei93};
18 = \cite{fer01};
19 = \cite{mar81};
20 = \cite{vie03};
21 = \cite{nis05};
22 = \cite{mey03};
23 = \cite{gra06};
24 = \cite{her14};
25 = \cite{joy54};
26 = \cite{sua06}.
}
\tablenotetext{d}{
    1 = Gaia DR2;
    2 = \cite{joh20};
    3 = \cite{tor06}.}
\tablenotetext{e}{
    2MASS = 2MASS Point Source Catalog;
    Fs18 = our FourStar photometry;
    VHS = VHS Data Release 6.
}
\tablenotetext{f}{
nodet = non-detection;
    sat = saturated;
    out = outside of the camera's field of view;
     bl = photometry may be affected by blending with a nearby star;
    bin = includes an unresolved binary companion;
    err = W2 magnitudes brighter than ~6 mag are erroneous;
  unres = too close to a brighter star to be detected;
    ext = photometry is known or suspected to be contaminated by
          extended emission (no data given when extended emission dominates);
  false = detection from WISE catalog appears false or unreliable based
          on visual inspection.}
\tablenotetext{g}{J-H and H-K = derived from these colors assuming
photospheric near-IR colors \citep{luh20};
IR spec = derived from an IR spectrum.}
\end{deluxetable}

\begin{deluxetable}{ccll}
\tabletypesize{\scriptsize}
\tablecaption{IRAC Observations of Corona Australis\label{tab:epochs}}
\tablehead{
\colhead{AOR} & \colhead{PID} & \colhead{PI} & \colhead{epoch}
}
\startdata
3650816&6&G. Fazio&2004.3\\
17672960&30784&G. Fazio&2006.7\\
17673472&30784&G. Fazio&2006.7\\
27041280&30574&L. Allen&2008.4\\
47018240&90071&A. Kraus&2012.9\\
47018496&90071&A. Kraus&2012.9\\
47018752&90071&A. Kraus&2012.9\\
47019008&90071&A. Kraus&2012.9\\
47019264&90071&A. Kraus&2012.9\\
47019520&90071&A. Kraus&2012.9\\
47019776&90071&A. Kraus&2012.9\\
47020032&90071&A. Kraus&2012.9\\
47020288&90071&A. Kraus&2012.9\\
47020544&90071&A. Kraus&2012.9
\enddata
\end{deluxetable}

\clearpage

\begin{deluxetable}{llll}
\tabletypesize{\scriptsize}
\tablewidth{0pt}
\tablecaption{Observing Log\label{tab:log}}
\tablehead{
\colhead{Telescope/Instrument} &
\colhead{Disperser/Aperture} &
\colhead{Wavelengths/Resolution} &
\colhead{Targets}}
\startdata
IRTF/SpeX & prism/$0\farcs8$ slit & 0.8--2.5~\micron/R=150 & 164 \\
4m Blanco CTIO/COSMOS & red VPH/$0\farcs9$ slit & 0.55--0.95~\micron/3~\AA & 148 \\
4m Blanco CTIO/ARCoIRIS & 110.5 l~mm$^{-1}$ + prism/$1\farcs1$ slit &0.8--2.47~\micron/R=3500 & 53 \\
Gemini South/GMOS & R400/$0\farcs75$ slit & 0.6--1~\micron/6~\AA & 1 \\
Gemini South/FLAMINGOS-2 & HK Grism/$0\farcs72$ slit &1.10--2.65~\micron/R=450 & 2 \\
Magellan/FIRE & prism/$0\farcs8$ slit & 0.8--2.5~\micron/R=100 & 23
\enddata
\end{deluxetable}

\begin{deluxetable}{lcccccc}
\tabletypesize{\scriptsize}
\tablewidth{0pt}
\tablecaption{Spectroscopic Data for Candidate Members of
Corona Australis\label{tab:spec}}
\tablehead{
\colhead{Source Name\tablenotemark{a}} &
\colhead{Spectral} &
\colhead{Instrument} &
\colhead{Date} &
\colhead{Young?} &
\colhead{Satisfies Gaia} &
\colhead{Adopted Member} \\
\colhead{} &
\colhead{Type} &
\colhead{} &
\colhead{} &
\colhead{} &
\colhead{Kinematic Criteria?} &
\colhead{of RCrA?}
}
\startdata
2MASS J18204248-3701412   &   M5     &   SpeX      &     2021 May 11     &       Y& Y& Y\\
2MASS J18205562-3426455   &   M8     &   SpeX      &     2020 Aug 14     &       Y& \nodata & N?\\
2MASS J18214722-3909565   &   M5.5   &   SpeX      &     2021 May 12     &       Y& Y& Y\\
2MASS J18215513-3718049   &   K7     &   COSMOS    &     2021 Jun 19     &       Y& Y& Y\\
2MASS J18222970-3434131   &   M5.5   &   COSMOS    &     2021 Jun 19     &       Y& Y& Y
\enddata
\tablecomments{This table is available in its entirety in a machine-readable form. A portion is shown is here for guidance regarding its form and content.}
\tablenotetext{a}{Identifications are from the 2MASS Point Source Catalog
when available.  Otherwise, they are from Gaia EDR3 or VHS DR6.}
\end{deluxetable}

\begin{deluxetable}{lccc}
\tabletypesize{\scriptsize}
\tablewidth{0pt}
\tablecaption{Gaia Candidates without Spectroscopy\label{tab:remain}}
\tablehead{
\colhead{Gaia EDR3} &
\colhead{Right Ascension\tablenotemark{a}} &
\colhead{Declination\tablenotemark{a}} &
\colhead{$G$} \\
}
\startdata
\cutinhead{Kinematic Candidates with RUWE$<$1.6}
4044760647651736064 & 275.630538 & $-$33.457568 & 14.11 \\
6733365816472640000 & 278.847121 & $-$36.348380 & 19.72\\
6733973361050980352 & 279.801448 & $-$34.870714 & 19.74\\
6729752954388414976 & 280.714523 & $-$38.492150 & 19.15\\
6733635914059722752 & 281.078293 & $-$35.640243 & 12.42\\
6729296107305930624 & 281.685646 & $-$38.872435 & 19.79\\
6730249800505083392 & 282.358207 & $-$37.277634 & 17.17\\
6732206067902438144 & 282.633911 & $-$35.201108 & 18.17\\
6731205138658643584 & 285.707301 & $-$36.773380 & 15.28\\ %12051499 (12051500)
6756775415446336640 & 286.206810 & $-$32.601102 & 16.91 \\
\cutinhead{Kinematic Candidates with RUWE$>$1.6}
6727177421432130560 & 277.322152 & $-$37.763763 & 14.06\\ %1888968 (1888966)
6735189150342584320 & 281.545001 & $-$34.944044 & 16.16\\ %8051890 (8051889)
\cutinhead{Non-kinematic Candidate Companions}
6734638707323220992 & 278.809732 & $-$34.718309 & 20.65\\ %7312799 (7312800)
6730058107518665344 & 279.848667 & $-$37.497411 & 19.51\\ %2300468 (2300467)
6733690309836940672 & 280.231271 & $-$35.371319 & 19.83\\ %6885325 (6885326)
6733691134465880192 & 280.553312 & $-$35.610333 & 20.52\\ %6853171 (6853170)
6730364218421820928 & 280.760560 & $-$37.461903 & 19.89\\ %2636127 (2636128)
6730572335360428416 & 281.160780 & $-$36.365688 & 20.37\\ %2767737 (2767736)
6730228360027665536 & 281.821826 & $-$37.567360 & 15.62\\ %2626584 (2626585)
6731791659404134528 & 283.032778 & $-$36.692146 & 13.93\\ %11365690 (11365690)
6717001475652511488 & 283.738020 & $-$39.854994 & 20.65\\ %10851468 (10851467)
6731220085144465280 & 285.419170 & $-$36.742231 & 14.68\\ %12050976 (12050975)
6719180123582696192 & 285.799478 & $-$37.150721 & 16.27
%11987830 (11987829)
\enddata
\tablenotetext{a}{From Gaia EDR3 (ICRS at Epoch 2016.0).}
\end{deluxetable}

\begin{deluxetable}{lcccc}
\tabletypesize{\scriptsize}
\tablewidth{0pt}
\tablecaption{Disk Fractions in Corona Australis\label{tab:diskfraction}}
\tablehead{
\colhead{Sample} &
\colhead{$<$K6} &
\colhead{K6--M3.5} &
\colhead{M3.75--M5.75} &
\colhead{M6--M8}}
\startdata
$b\geq-17.3\arcdeg$ and $A_J\leq1$ & 2/11=$0.18^{+0.16}_{-0.06}$ & 9/68=$0.13^{+0.05}_{-0.03}$ & 18/120=$0.15^{+0.04}_{-0.03}$ & 10/39=$0.25^{+0.08}_{-0.06}$ \\
$b<-17.3\arcdeg$ and $A_J\leq1$ & 1/6=$0.17^{+0.23}_{-0.06}$ & 6/17=$0.35^{+0.13}_{-0.09}$     & 12/34=$0.35^{+0.09}_{-0.08}$  & 5/7=$0.71^{+0.10}_{-0.20}$\\
$b<-17.3\arcdeg$ and $A_J>1$    & 3/3=$1.00^{+0.00}_{-0.37}$ & 7/9=$0.77^{+0.08}_{-0.18}$   & 4/9=$0.44^{+0.16}_{-0.14}$ & 1/3=$0.33^{+0.28}_{-0.15}$
\enddata
\end{deluxetable}

\clearpage

\begin{figure}[h]
  \centering
  \includegraphics[trim = 0mm 0mm 0mm 0mm, clip=true, scale=.9]{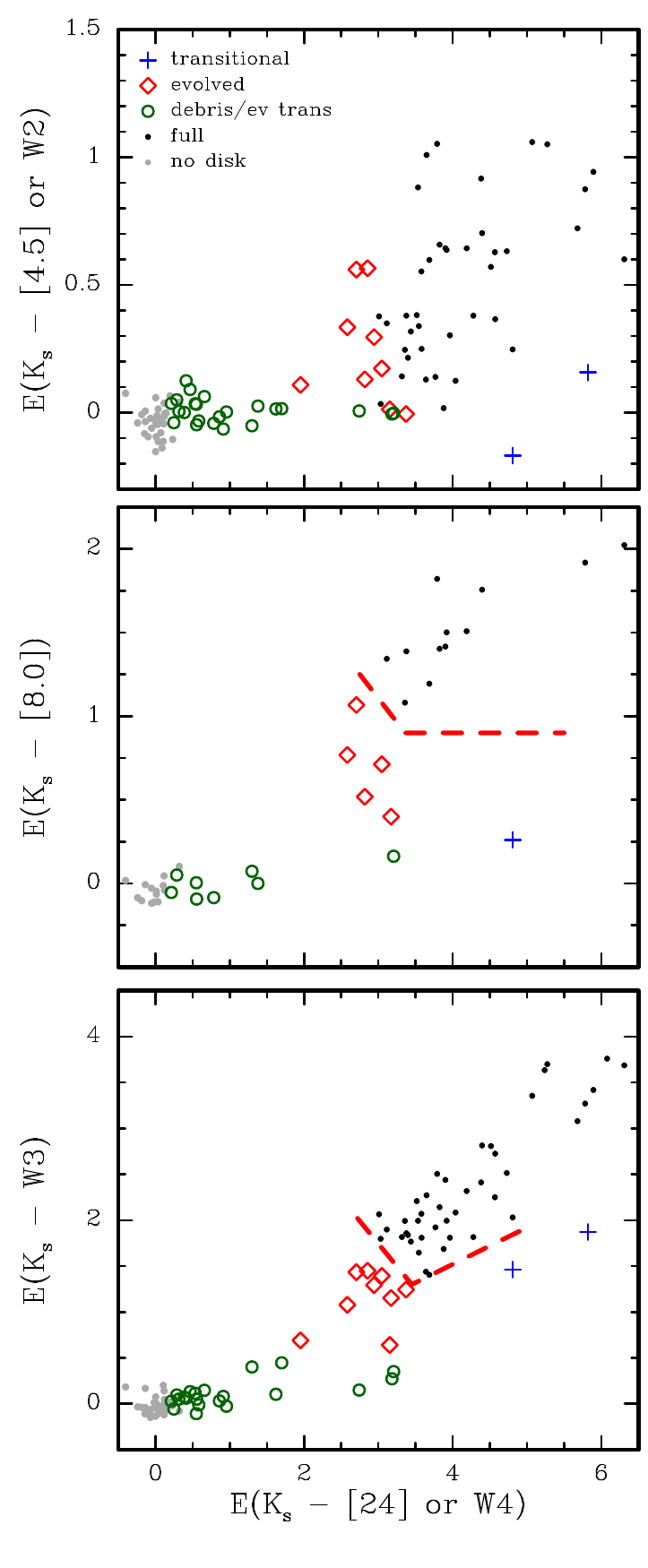}
\caption{
Extinction-corrected IR color excesses for adopted members of Corona Australis,
If a source lacked [4.5] or [24] photometry, the W2 or W4 data is shown instead.
The bottom two diagrams include boundaries that are used to distinguish full 
disks from disks in more advanced stages of evolution (dashed lines).
}
\label{fig:diskclass}
\end{figure}

\end{document}